%% file: lmetric.tex
\documentclass[10pt,sigplan,letterpaper,twocolumn]{article}
\usepackage[10pt,nocopyright]{sigmin}

\usepackage{listings}
\usepackage{color}
\usepackage{dingbat}
\usepackage[short]{optidef} %

\usepackage{epsfig,makecell}                                      %
\usepackage[hyphens]{url}                                         %
\usepackage[breaklinks,colorlinks]{hyperref}                      %
\hypersetup{citecolor=blue,linkcolor=blue}                        %
\usepackage{amsmath,amsopn,amssymb,amsthm}                        %
\usepackage{thmtools}
\usepackage{thmbox}
\usepackage[toc,page]{appendix}
\usepackage{subcaption}                                           %
\usepackage{endnotes,microtype,xspace,fancyvrb,multirow} %
\usepackage{epigraph} 
\usepackage{stfloats}

\usepackage{booktabs}      
\usepackage{tabularx}
\usepackage{array,underscore,relsize}                             %
\usepackage[T1]{fontenc}                                          %
\usepackage{times}                                                %
\usepackage{fancyhdr,lastpage}                                    %
\usepackage{enumitem}                                             %
\usepackage[labelfont=bf,font=normal,skip=1pt]{caption}           %
\usepackage[export]{adjustbox}                                    %
\newcommand{\burl}{\url}                                             %
\usepackage{pifont}                                               %
\usepackage{tablefootnote}                                        %
\usepackage{bbding}                                        %
\usepackage[table]{xcolor} %
\usepackage[linesnumbered,vlined,boxed,commentsnumbered,ruled]{algorithm2e}
\usepackage{float}
\DontPrintSemicolon

\newcommand{\nospacestitle}[1]{\noindent{\bf #1}}

\usepackage{amssymb}
\usepackage[normalem]{ulem}
\usepackage[hang,flushmargin]{footmisc}
\usepackage{textcomp}
\usepackage{graphicx}
\usepackage{authblk}
\usepackage[available]{usenixbadges}  

\usepackage[compact]{titlesec}
\titleformat*{\section}{\large\bfseries}
\titleformat*{\subsection}{\normalsize\bfseries}
\titleformat*{\subsubsection}{\normalsize\bfseries}
\titlespacing{\section}{0pt}{3ex}{1ex}
\titlespacing{\subsection}{0pt}{2ex}{1ex}

\hypersetup{
  colorlinks,
  linkcolor={red!50!black},
  citecolor={blue!50!black},
  urlcolor={blue!80!black}
}

\newcommand{\sys}{{\textsc{LMetric}}} %
\newcommand{\kvcache}{\textsc{KV\$}}

\newcommand{\company}{\textsc{{Bailian}}}

\newcommand{\fig}[1]{Figure{~\ref{#1}}}

\newcommand{\one}{\texttt{\uppercase\expandafter{\romannumeral1}}}
\newcommand{\two}{\texttt{\uppercase\expandafter{\romannumeral2}}}

\newcommand{\stitle}[1]{\vspace{1.1ex}\noindent{\bf #1}}

\newcommand{\metric}[1]{\textbf{\uline{#1}}}

\usepackage{tikz}
\usepackage{colortbl}
\usetikzlibrary{tikzmark,calc}

\usepackage{balance}
\newcolumntype{P}[1]{>{\centering\arraybackslash}p{#1}}

\definecolor{appcolor}{RGB}{191,255,255}

\makeatletter
\renewcommand\AB@affilsepx{ \quad\protect\Affilfont \, } %
\makeatother

\SetKwInput{KwInput}{Input}
\SetKwInput{KwOutput}{Output}
\SetKwComment{Comment}{$\triangleright$\ }{}

\begin{document}

\title{\Large \bf{Simple is Better: Multiplication May Be All You Need for LLM Request Scheduling}}

\setlength{\affilsep}{0.3em}
\renewcommand\Authsep{,\quad}
\renewcommand\Authand{,\quad}
\renewcommand\Authands{,\quad}
\author[1]{Dingyan Zhang\textsuperscript{\dag}}
\author[1]{Jinbo Han\textsuperscript{\dag,\ddag}}
\author[1]{Kaixi Zhang\textsuperscript{\dag,\ddag}}
\author[1]{Xingda Wei\,{\Envelope}}
\author[2]{Sijie Shen}
\author[2]{Chenguang Fang}
\author[2]{Wenyuan Yu}
\author[2]{Jingren Zhou}
\author[1]{Rong Chen}

\affil[1]{\vspace{-2.mm}Institute of Parallel and Distributed Systems, Shanghai Jiao Tong University}
\affil[2]{Alibaba Group\vspace{-1.mm}}
\date{}
\maketitle

\begin{NoHyper}
\def\thefootnote{*}\footnotetext{Our system name \sys{} stands for \textbf{L}arge \textbf{M}odel metric, paying homage to Lyapunov's stability theory and Markov's queueing theory.}
\def\thefootnote{\dag}\footnotetext{Most work done when intern at Alibaba Group.}
\def\thefootnote{\ddag}\footnotetext{Jinbo Han and Kaixi Zhang contributed equally and significantly to this work.}
\def\thefootnote{\Envelope}\footnotetext{Xingda Wei is the corresponding author (\url{wxdwfc@sjtu.edu.cn}).}
\end{NoHyper}

\renewcommand{\thefootnote}{\arabic{footnote}}

\frenchspacing

\vspace{2em}

\input{abs}
\input{intro-v2}
\input{bg}

\input{factory}

\input{principles}

\input{method-v1.tex}

\input{eval.tex}

\input{dislim.tex}

\input{related.tex}

\input{concl}

\input{ack}

\balance

{\small
\bibliographystyle{acm}
\bibliography{lmetric}
}

\clearpage
\appendix
\twocolumn
\input{appendix-v1.tex}

\clearpage

\end{document}

%% file: abs.tex
\begin{abstract}
    \noindent
High-quality LLM request scheduling requires
meeting two key objectives: ensuring the routed instance has
{\kvcache} to accelerate request execution, and ensuring
that the workload is balanced across instances.
Achieving both objectives is challenging because
pursuing one may compromise the other.
Current approaches use various combinators (e.g., linear combinations)
to compute a scheduling score that combines indicators for the two objectives.
These approaches are complex: they either require significant workload-specific hyperparameter tuning
or model-hardware-aware simulator development,
yet could still lead to suboptimal performance.

In this paper, we show that using a simple multiplication of two carefully chosen indicators---one
{\kvcache}-aware (new prefill tokens if routed to an instance) and one load-balancing-aware
(current batch size of the instance)---as the scheduling score ({\sys}\textsuperscript{*})
can achieve both objectives simultaneously without any hyperparameter tuning.
The key idea is that the simply multiplied score considers both objectives in a manner similar to a linear combination,
but the original hyperparameters cancel out during comparison,
so no tuning is needed to find the best parameters.
The two indicators are chosen based on our analysis of LLM characteristics.
Our extensive experiments show that this simple approach can reduce TTFT by 92\% and 39\%, and TPOT by 24\% and 51\%,
compared to vLLM-v1 and an in-production scheduler on real-world workloads covering chatbots 
and coding agents.
We also derive the mathematical conditions under which multiplication may fail,
and find that such conditions are extremely rare in practice and
can be detected (and mitigated) beforehand.

{\sys} has been deployed in production
and canary release confirms its effectiveness. 

\end{abstract}

%% file: intro-v2.tex
\section{Introduction}
\label{sec:intro}

\noindent
This paper studies how to efficiently route LLM requests to a cluster 
of serving instances---the minimal LLM engine deployment unit.
Serving LLMs has become a key building block in modern society,
and LLM providers typically deploy clusters of instances for serving,
where each cluster contains a \emph{global scheduler} that routes incoming requests to the instances it manages~\cite{305212,DBLP:conf/sosp/Xiang0QYZYZL0025,
DBLP:journals/corr/abs-2505-09999, llm-d, ai-Dynamo, aigw, vllm-code}.
Upon receiving a request, the instance generates result tokens in two phases:
The prefill (P) phase generates the first result token, and its serving quality is measured by time-to-first-token (TTFT).
The decode (D) phase then generates the remaining tokens in a streaming manner, and its serving quality is measured by time-per-output-token (TPOT).

Providing an effective scheduling policy is crucial for cluster-level LLM serving.
This is because, 
similar to traditional request routing~\cite{DBLP:conf/asplos/DelimitrouK14,DBLP:conf/eurosys/ZhangTHJGW13,DBLP:conf/osdi/GogSGWH16},
better request placement significantly reduces overall latency (lower TTFT and TPOT)
due to factors such as improved load balancing across instances. 
Low-latency serving is especially critical for current
interactive applications such as ChatGPT~\cite{chatgpt} and copilots~\cite{copilot}, as it is essential to meeting user expectations~\cite{latency-study,DBLP:conf/osdi/ZhongLCHZL0024,aws-latency}.
Moreover, recent agentic workloads consume tokens rapidly through interactive programs~\cite{claudeapi,geminiapi,openaiapi}.

Achieving a good LLM-specific scheduling policy is non-trivial:
First, considering only load balancing across instances---which is adopted by 
a recent state-of-the-art serving system vLLM~\cite{vllm-code} and
traditional request routing---%
is insufficient.
This is because the computation required to process each request differs across
instances due to {\kvcache},
the intermediate context of the processed tokens (\textsection{\ref{sec:kvcache-aware}}).
Specifically,
if an incoming request's (partial) input tokens hit the {\kvcache} cached on an instance,
the instance can skip generating the corresponding {\kvcache} for the hit tokens,
thereby accelerating the subsequent prefill and decode phases.
However,
incorporating only {\kvcache}-aware indicators into scheduling decisions 
(e.g., the {\kvcache} hit ratio if routing a request to an instance)
is also insufficient,
because it biases requests towards instances with {\kvcache} hits
and hurts load balancing across instances 
 (\textsection{\ref{sec:challenge-balancing}}).

To balance the two objectives---{\kvcache}-awareness and load balancing---three combination strategies exist. 

First, the linear combination strategy (i.e., weighted sum)~\cite{vllm,ai-Dynamo} combines the indicators for each objective---per-instance metrics that score each objective if the request were routed to that instance, e.g., the {\kvcache} hit ratio for {\kvcache}-awareness and the current batch size for load balancing---into a single score for scheduling (\textsection{\ref{sec:linear-combination}}).
It is a popular choice adopted by current works 
and one of the world's leading LLM service providers (Alibaba {\company}).
However, linear combination requires complex workload-specific hyperparameter tuning to achieve both objectives.
Moreover, a statically tuned hyperparameter is suboptimal because workloads may change dynamically (\textsection{\ref{sec:linear-combination}}).

Second, the filter-based strategy first filters out instances that are suspected to suffer from imbalanced workloads,
and then selects the instance with the most {\kvcache} hits among the remaining instances~\cite{aibrix}.
It still requires non-trivial workload-specific tuning to determine the filtering threshold.
Worse still, 
its scheduling is biased towards load balancing and
thus cannot fully utilize the {\kvcache} (\textsection{\ref{sec:filter-combination}}).

Finally, the simulation-based strategy~\cite{305212, llm-d, DBLP:journals/corr/abs-2507-17769, DBLP:journals/corr/abs-2504-08784}
first uses a simulator to predict the expected latency of routing a request to each instance,
and then takes the latency as the scheduling score.
The simulator estimates the latency based on its current indicators, e.g., stored {\kvcache} and request load in progress (queued requests),
so the latency score can be viewed as 
a high-order combination of the {\kvcache} and load-balance indicators 
to achieve the best of both worlds.
However, the effectiveness of the strategy relies heavily on the accuracy of the simulator,
which requires complex per-model, per-hardware, and per-deployment development.
Otherwise,
an inaccurate simulation can lead to poor scheduling performance (\textsection{\ref{sec:prediction-combination}}).
Even with an accurate simulator, it may still fail to achieve performance comparable
to other candidates in certain cases.

In this paper, we show that using the multiplication of one indicator for {\kvcache}-awareness
and one indicator for load balancing as the scheduling score
can effectively combine the two objectives
without complex hyperparameter tuning or any simulator.
The key idea is to replace the addition operation in a linear combination
with multiplication.
The resulting score preserves a trend similar to that of a linear combination,
and the hyperparameters cancel out during score comparisons among instances,
so no tuning is required (\textsection{\ref{sec:method}}).

Making this simple method work well in practice requires care.
First,
we found that careful indicator selection can further improve the method's effectiveness (\textsection{\ref{sec:indicators}}).
For example,
using the number of queued new prefill tokens when routing a request to an instance, considering {\kvcache} hits,
as the {\kvcache}-aware indicator is
better than using the {\kvcache} hit ratio.
Second, 
we mathematically derive the approximate conditions under which multiplication may fail.
Based on the formulated conditions, we found that they are extremely rare in practice (\textsection{\ref{sec:m-analyze}})---occurring under
extreme {\kvcache} skewness that compromises load balance.
Hence, we further design a
two-phase approach to detect and mitigate it.
Upon detection, we can fall back to a load-balancing-only policy.

We have compared our method with state-of-the-art systems, including vLLM~\cite{vllm-code}, ai-Dynamo~\cite{ai-Dynamo}, llm-d~\cite{llm-d}, and
the one used in Alibaba {\company}~\cite{bailian-product} on real LLM serving workloads
covering chatbots, API calling, and coding agents (\textsection{\ref{sec:eval}}).
On an H20 cluster with up to 16 GPUs, evaluation on popular models covering both dense and MoE architectures
confirms the benefits of our approach.
{\sys} has been deployed in production at {\company} on hundreds of GPUs, and 
performance observed from a canary release confirms its effectiveness.

\noindent\begin{minipage}{\columnwidth}
\stitle{Contributions. \,} This paper makes three contributions:
\begin{itemize}[leftmargin=*, topsep=2pt, itemsep=2pt, parsep=0pt]
\item The first systematic study of how to efficiently schedule LLM serving requests in a cluster (\textsection{\ref{sec:principles}}).
\item The first multiplicative combination for efficient LLM request scheduling (\textsection{\ref{sec:method}}).
\item Extensive analysis and evaluation that confirm the effectiveness of the multiplicative approach (\textsection{\ref{sec:indicators}}, \textsection{\ref{sec:m-analyze}}, \textsection{\ref{sec:eval}}).
\end{itemize}
\end{minipage}

\vspace{3pt} Our code is open-sourced at \burl{https://github.com/blitz-serving/blitz-router},
and all our traces can be found at \burl{https://github.com/alibaba-edu/qwen-bailian-usagetraces-anon}. 

\stitle{Discussion: PD-colocation vs. PD-disaggregation. \,}
We focus on PD-colocated serving in this paper,
where both prefill and decode requests are served on the same instance. 
While there also exist deployments where prefill and decode requests are served on different instances (PD-disaggregation)~\cite{DBLP:conf/isca/PatelCZSGMB24,DBLP:conf/osdi/ZhongLCHZL0024,DBLP:journals/corr/abs-2401-11181},
PD-colocated serving is still widely adopted in practice~\cite{DBLP:conf/sosp/Xiang0QYZYZL0025}
because it is easier to maintain (no instance role management),
does not rely on fast networking between instances,
and yields better performance under certain conditions~\cite{revisit-disaggregated,tokenscale}.

We discuss how our observations and solutions apply to PD-disaggregation in \textsection{\ref{sec:dislim}}.

%% file: bg.tex
\section{LLM Serving and Scheduling}
\label{sec:bg}

\label{sec:llm-scheduling}

\begin{figure}[!t]
        \vspace{2mm}
        \begin{minipage}{1\linewidth}
        \centering
        \includegraphics[width=0.92\linewidth, trim=0.25cm 23.92cm 43cm 0.25cm, clip]{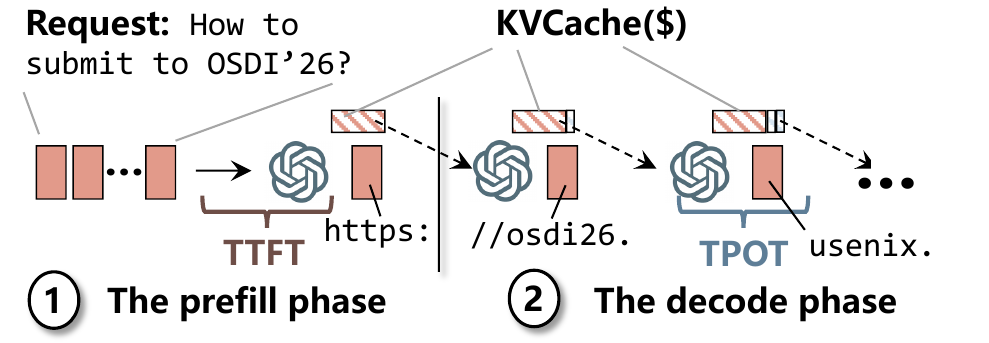}
        \end{minipage} \\[2pt]
        \begin{minipage}{1\linewidth}
        \caption{\small{%
            An illustration of how an LLM generates output tokens,
            and the two performance metrics: 
            time-to-first-token (TTFT) and time-per-output-token (TPOT).
        }}
        \label{fig:llm-basics}
        \end{minipage} \\[-15pt]
        \end{figure}

\nospacestitle{Serving requests with an LLM ({\fig{fig:llm-basics}}).   \,}
LLMs generate tokens auto-regressively in two steps:
\ding{192} In the \emph{prefill} phase, the input tokens are fed into the model to
produce the first output token.
The LLM then enters the \emph{decode} phase (\ding{193}) to generate subsequent tokens one-by-one,
each conditioned on all prior input and generated tokens.
To accelerate decode, the \emph{processed context} is materialized as tensors in GPU memory,
termed the \emph{key-value cache} ({\kvcache}).
Consequently, decode computation is far smaller than prefill, since the {\kvcache} for input tokens is already available and not regenerated.

Two key serving-quality metrics are
\emph{time-to-first-token (TTFT)} and \emph{time-per-output-token (TPOT)}:
TTFT determines the service's responsiveness to user requests,
while TPOT affects both subsequent responsiveness and overall request completion time.

\begin{figure}[!t]
        \hspace{3mm}
        \begin{minipage}{1\linewidth}
\includegraphics[width=0.92\linewidth, trim=0.25cm 22.9cm 46cm 0.25cm, clip]{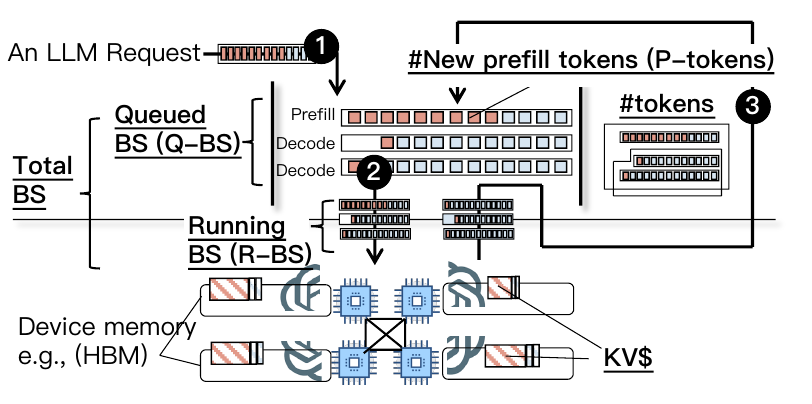}
        \end{minipage} \\[4pt]
        \begin{minipage}{1\linewidth}
        \caption{\small{%
            The system view of how an LLM serving instance handles
            requests and representative direct system \metric{indicators}
            that can be collected by the global scheduler. %
            The detailed meaning of each indicator will be described at first use. 
            \textbf{BS} is the abbreviation for \uline{b}atch \uline{s}ize.
        }}
        \label{fig:llm-instance}
        \end{minipage} \\[-12pt]
        \end{figure}

\stitle{Serving instance and {\kvcache} cache ({\fig{fig:llm-instance}}). \,}
An instance is the minimum unit that serves requests and hosts one complete copy of the LLM's parameters,
potentially spanning multiple GPUs if the model exceeds a single GPU's memory.
The serving flow is the same in either case:
an incoming request is pushed into a queue (prefill or decode) on the instance (\ding{182}).
Once the GPU(s) become available, queued requests are batched and
executed efficiently with \emph{chunked prefill}~\cite{298679} (\ding{183}).
The instance then re-enqueues any request that needs further decoding (\ding{184}).

Serving is \emph{stateful}: a request's {\kvcache}
is cached in GPU or CPU memory even after generation finishes ({\kvcache} cache~\cite{305212,298501,10.5555/3768039.3768067}).
If a future request routed to the instance shares a prefix with a cached {\kvcache},
the instance can skip computing the matched prefix tokens, significantly reducing computation cost.
In {\fig{fig:llm-instance}}, blue tokens skip computation on a {\kvcache} hit.

\begin{figure}[!t]
        \begin{minipage}{1\linewidth}
        \centering
        \includegraphics[width=0.85\linewidth, trim=0.25cm 15.7cm 38.7cm 0.25cm, clip]{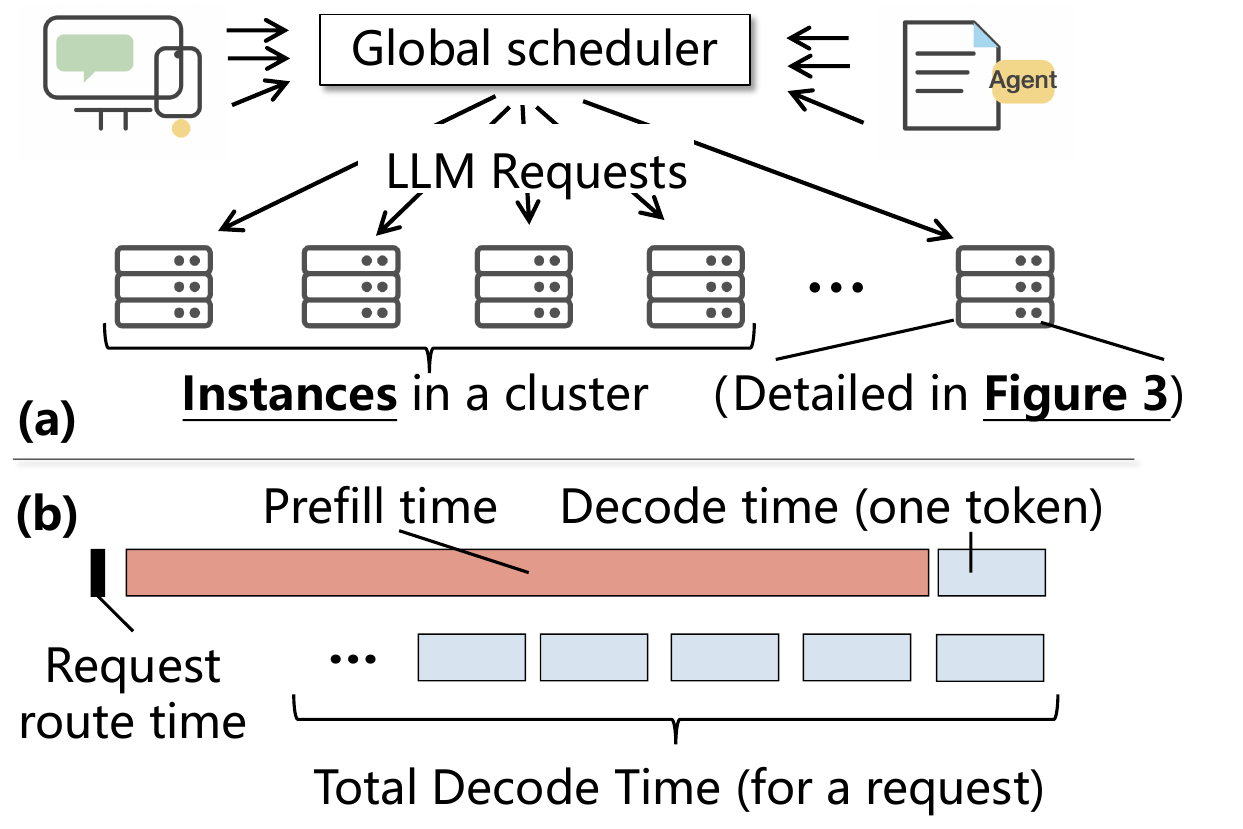}
        \end{minipage} \\[4pt]
        \begin{minipage}{1\linewidth}
        \caption{\small{%
        (a) System view of an LLM serving cluster,
        and
        (b) Comparison of per-request serving time and routing cost.
        }}
        \label{fig:setup}
        \end{minipage} \\[-15pt]
        \end{figure}

\stitle{Request scheduling in a cluster ({\fig{fig:setup}}). \,}
To handle large request volumes,
providers deploy clusters of instances, each with a dedicated \emph{global scheduler}
that routes requests to its managed instances, as shown in (a).
The scheduler runs a \emph{scheduling policy} to pick a destination instance per request---our focus.
Providers may deploy multiple clusters of the same model for reliability and geo-affinity,
but inter-cluster routing is out of the scope of this paper.

All existing scheduling policies follow a three-step process:
the router first (optionally) \emph{filters} instances,
then \emph{scores} the remainder by a preference order,
and routes the request to the best-scoring one.
Scores derive from per-instance indicators, described in \textsection{\ref{sec:principles}}.

%% file: factory.tex
\section{The Analysis Framework}
\label{sec:factory}

\noindent
To analyze different scheduling policies in an apples-to-apples manner, we implement
a flexible LLM scheduling analysis framework.
Two drivers shape its design.
(1) Existing policies are buried in concrete open- and closed-source serving-system
implementations, making apples-to-apples comparison hard.
For example, AIBrix's~\cite{aibrix} Go reimplementation of vLLM's policy runs
6.2\,$\times$ faster than vLLM's Python version~\cite{vllm-code} due to a later-confirmed
performance bug~\cite{vllm-bug};
our Rust implementation is a further 1.2\,$\times$ faster than AIBrix on the same policy.
(2) Though complex, all existing policies boil down to computing scheduling scores from
per-instance indicators.
A unified indicator factory thus lets developers implement and explore new policies in
a few lines of code (described below).

\begin{figure}[!t]
        \begin{minipage}{1\linewidth}
        \centering
        \includegraphics[width=0.95\linewidth, trim=0.25cm 14.75cm 33cm 0.25cm, clip]{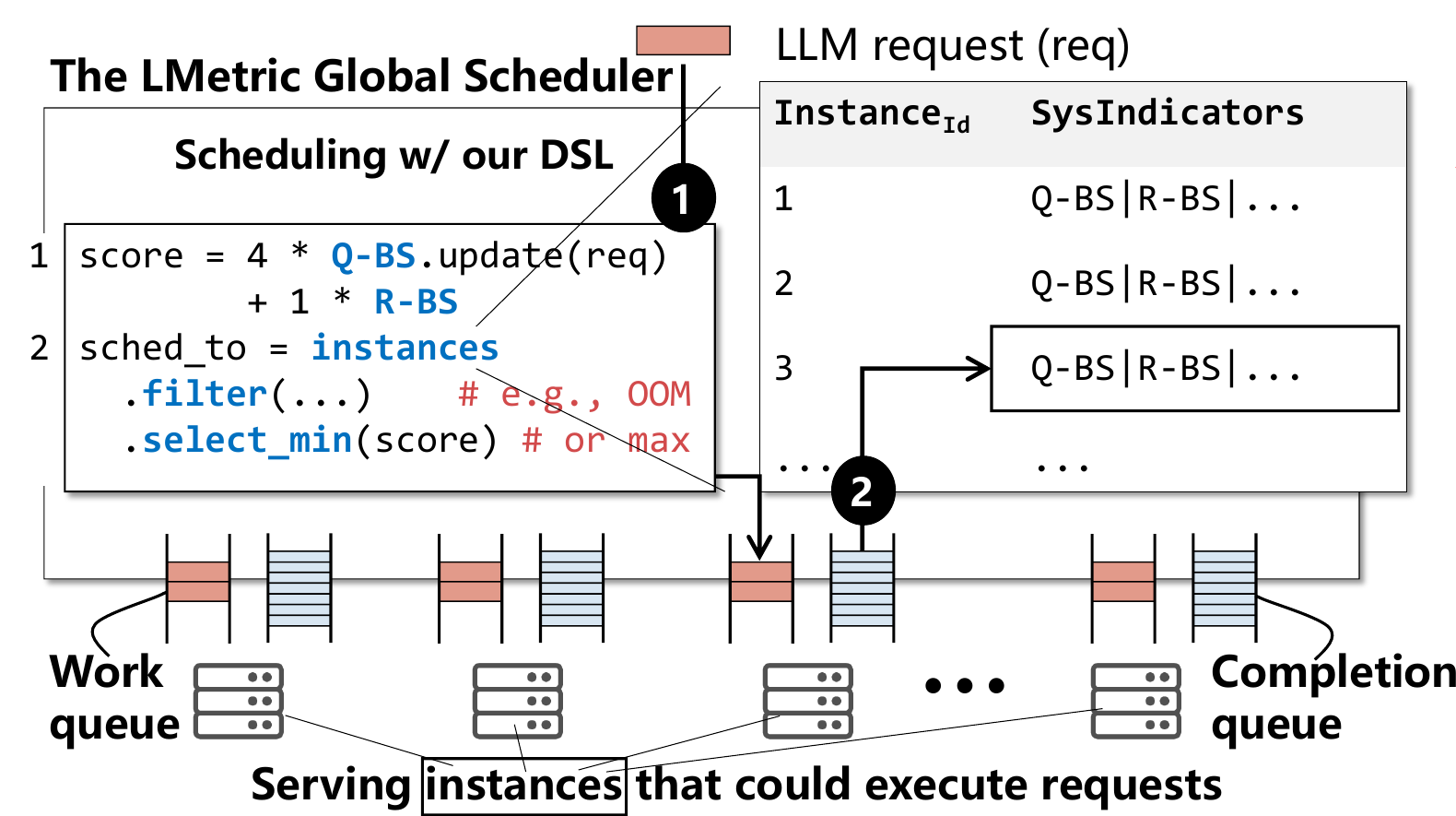}
        \end{minipage} \\[5pt]
        \begin{minipage}{1\linewidth}
        \caption{\small{%
        The system architecture of our indicator factory
        and its programming model for scheduling policies.
        }}
        \label{fig:factory}
        \end{minipage} \\[-10pt]
        \end{figure}

\stitle{Indicator factory. \,}
Our framework is a standalone inference router written in Rust that works with any LLM
serving engine; Rust keeps it efficient and robust.
Its core component, shown in {\fig{fig:factory}}, is an \emph{indicator factory} that
automatically collects and, when needed, computes the indicators each scheduling policy
requires.
For scalability, the factory piggybacks indicator collection on instance responses:
the router maintains a long-lived connection to each engine instance and, on every
response, extracts the required indicators from the response header and updates the
factory.

\stitle{Programming model.  \,}
Our framework provides a simple API to implement different scheduling policies.
The API lets developers express scheduling as a function over the factory's symbolic
per-instance indicators in a few lines.
On top of the score function, we further provide primitives to filter instances and
to select the instance with the minimum (or maximum) score.
Line 1 of {\fig{fig:factory}} shows the policy adopted by vLLM~\cite{vllm-code} using
our framework: the score is a weighted sum of \texttt{Q-BS} (queued batch size---the
number of queued requests in an instance's queue) and \texttt{R-BS} (the number of
running requests within batch).
On a new request (line 4), the router retrieves and computes the indicator values from
the factory in parallel, derives a score for every instance, and routes the request to
the instance with the minimum score.

%% file: principles.tex
\section{Characterizing LLM Request Scheduling}
\label{sec:principles}

\begin{figure*}[!t]
    \hspace{1mm}
     \includegraphics[width=0.99\linewidth, trim=0.25cm 22.45cm 36cm 0.25cm, clip]{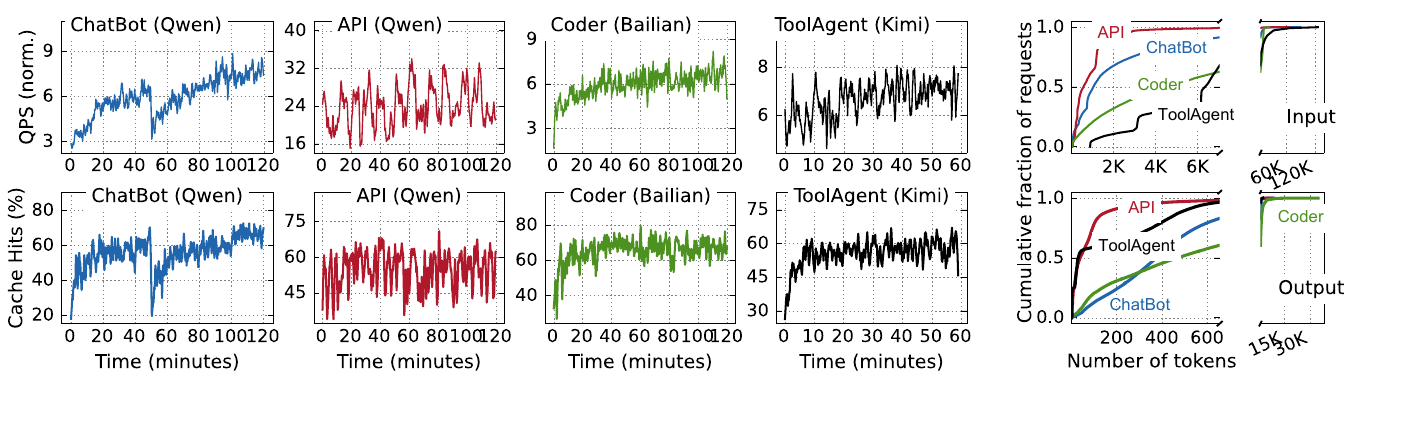} \\[-20pt]
    \begin{minipage}{1\linewidth}
    \caption{\small{
       Our studied traces that cover major scenarios in powering LLM services.
        }}
    \label{fig:traces}
    \end{minipage} \\[-20pt]
\end{figure*}

\subsection{Characterization Methodology}
\label{sec:eval-setup}

\nospacestitle{Testbed and instance used. \,}
Unless otherwise specified, all experiments run on a testbed with 16 NVIDIA H20 GPUs---hardware similar to that used for hosting LLM services at {\company}.
Each GPU has 96\,GB HBM, sufficient for common models available on the market~\cite{DBLP:conf/sosp/Xiang0QYZYZL0025}.
The router runs on a high-end CPU server with 160 Intel Xeon cores and 1\,TB of DRAM.
Our instance runs vLLM-v1 (vLLM)~\cite{vllm-code}---the state-of-the-art LLM serving engine
with the latest optimizations, such as chunked prefill~\cite{298679} and fast GPU kernels~\cite{flashinfer}.

\stitle{Models. \,} 
We chose LLM models representative of different architectures and popular market choices,
including dense (Qwen2-7B) and mixture-of-experts (MoE) models (Qwen3-30B)~\cite{DBLP:conf/sosp/Xiang0QYZYZL0025}. 

\stitle{Workloads. \,}
We analyze real-world LLM serving traces, both open-source and collected from {\company}, covering common LLM applications:
\textbf{ChatBot (Qwen)} and \textbf{Agent (Qwen)}~\cite{bailian}
are open-sourced Alibaba Cloud traces that collect requests from a chatbot service similar to ChatGPT and
an LLM API calling agent service~\cite{claudeapi,geminiapi,openaiapi}, respectively.
\textbf{Coder} collects requests issued by coding agent services to a dedicated cluster in {\company} on a single day in November 2025,
and \textbf{ToolAgent (Kimi)}~\cite{mooncake-trace} is another open-sourced trace from Kimi
that collects requests from an agent service.
All traces except ToolAgent (Kimi) are collected from a single cluster routed by one global router;
the source of ToolAgent (Kimi) does not specify whether it is collected the same way.

Our analyzed workloads are representative: they span broad application scenarios,
and each selected trace preserves the essential characteristics for evaluating LLM scheduling policies.
Specifically, all requests in our traces contain the (hashed) content and request-issuance timestamp,
which are critical for evaluating the impact of {\kvcache}-aware scheduling on global scheduling (see \textsection{\ref{sec:kvcache-aware}}).
Other popular datasets like AzureLLMTrace~\cite{AzurePublicDataset} or BurstGPT~\cite{wang2025burstgpt} do not provide
such content.
Note that with traces containing hashed content, we can still replay them
with behavior that exactly matches the original~\cite{kvcache}.
To facilitate future research,
our high-performance trace replayer is open-sourced~\cite{blitzTraceReplayer}.

{\fig{fig:traces}} visualizes key features of our evaluated traces,
including their request arrival rates, input and output token numbers,
and the {\kvcache} hit rates assuming an infinite {\kvcache} space.
The request arrival rate is normalized due to confidentiality considerations
required for the \textbf{Coder} trace. 
Across all traces, over a given serving interval (e.g., 1 hour),
the request arrival and {\kvcache} hit rates are relatively stable, with a few short-term fluctuations.
Input and output token numbers vary across traces but are typically modest except for a few outliers.

\stitle{Trace scaling. \,}
Since the traces are collected from clusters at different scales than our testbed,
we scale them according to our testbed capability
similar to prior work~\cite{DBLP:conf/asplos/MiaoSDXL0J24,DBLP:journals/pvldb/AliPYS22,DBLP:conf/osdi/GujaratiKAHKVM20,DBLP:conf/osdi/ZhangWLWS0025,traceupscaler}.
Unless otherwise specified, we scale the average request arrival rate to
half of the maximum rate of our testbed obtained via offline profiling.
This approximates the serving configurations in {\company} because
when the arrival rate approaches serving capacity,
{\company} commonly reroutes requests to another underloaded cluster
or simply rejects them in non-critical cases (e.g., ChatApp)~\cite{305212}.
Otherwise, the service-level objectives (SLOs) of many requests cannot be met due to queuing~\cite{queuing}.
Our end-to-end analysis in \textsection{\ref{sec:eval}} further measures
the impact of different request arrival rates on scheduling performance,
and shows consistent results.

\subsection{Load-balancing Alone is Insufficient for LLM}
\label{sec:kvcache-aware}

\begin{figure}[!t]
    \centering
     \includegraphics[width=0.9\linewidth, trim=0.25cm 16.4cm 37.65cm 0.25cm, clip]{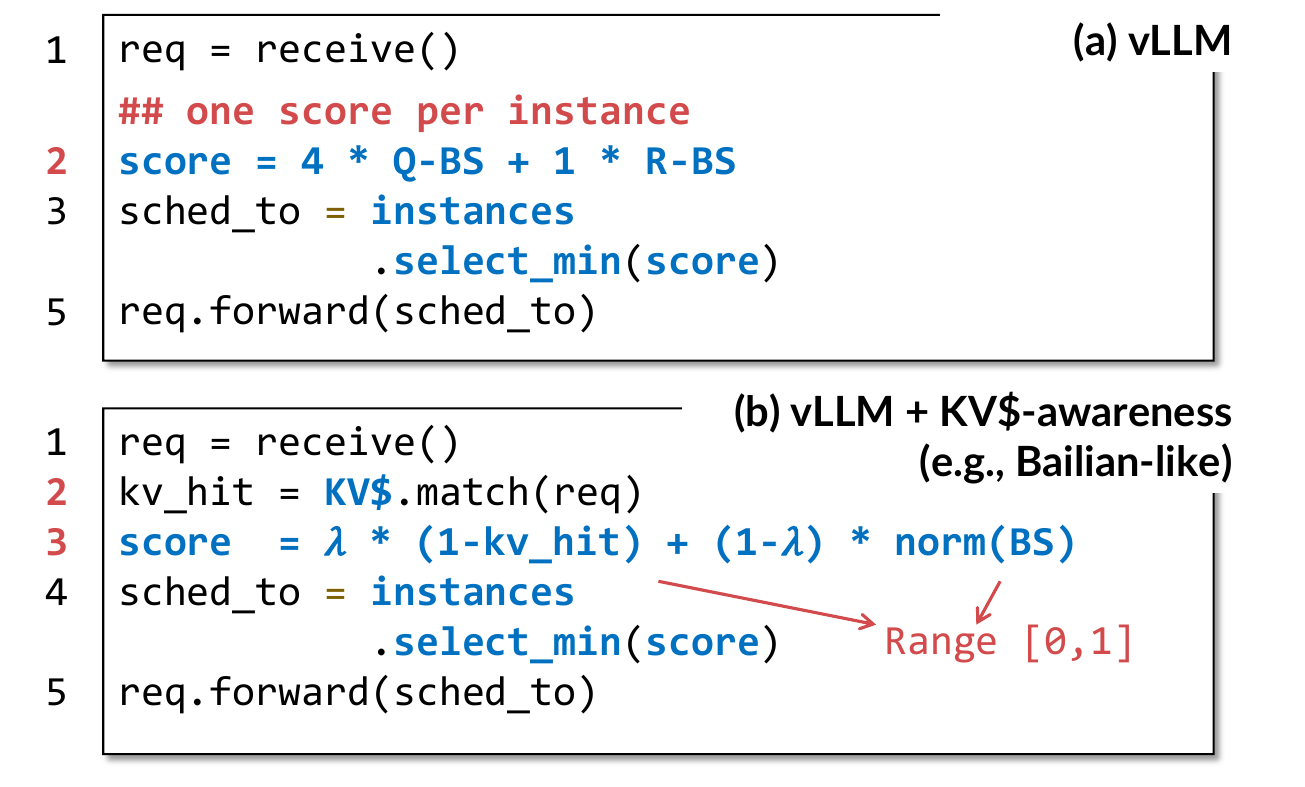} \\[5pt]
    \begin{minipage}{1\linewidth}
    \caption{\small{
        (a) The scheduling score of vLLM and 
        (b) adding {\kvcache}-awareness to its score for LLM scheduling. 
        Note that the linear combination of two indicators has only one degree of freedom (the weight $\lambda$).
        }}
    \label{fig:vllm-kvcache}
    \end{minipage} \\[-16pt]
\end{figure}

\begin{figure}[!t]
    \centering
    \includegraphics[width=1.0\linewidth]{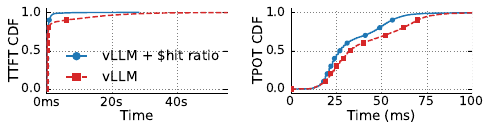} \\[-0pt]
    \begin{minipage}{1\linewidth}
    \caption{\small{   
        A comparison of the performance of vLLM and {\kvcache}-aware scheduling on ChatBot Trace with Qwen3-30B model.
        Other workloads and models are similar.
    }}
    \label{fig:kvcache-aware-cdf-1}
    \end{minipage} \\[-10pt]
\end{figure} 

\begin{figure}[t]
    \centering
    \includegraphics[width=0.97\linewidth]{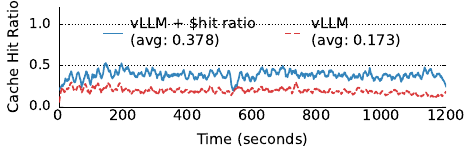} \\[0pt]
    \begin{minipage}{1\linewidth}
    \caption{\small{A {\kvcache} hit ratio comparison of vLLM vs. {\kvcache}-aware scheduling on ChatBot Trace with Qwen3-30B model.
    Other workloads and models are similar.
    }}    
    \label{fig:kvcache-aware-hit-ratio1}
     \end{minipage} \\[-11pt]
\end{figure}

\begin{figure}[t]
    \centering
    \includegraphics[width=0.97\linewidth]{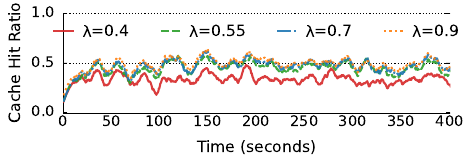} \\[1pt]
     \begin{minipage}{1\linewidth}
    \caption{\small{
A comparison of the {\kvcache} hit ratio by changing the weight of {\kvcache}-awareness
in the policy described in {\fig{fig:vllm-kvcache}} (b) on ChatBot Trace with Qwen3-30B model.     
    }}
    \label{fig:chln-cache-hit}
    \end{minipage} \\[-6pt]
\end{figure}

\begin{figure}[!ht]
    \centering
    \includegraphics[width=0.98\linewidth]{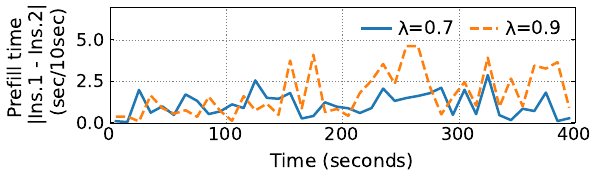} \\[-2pt]
    \caption{\small{
    A profile of the workload imbalance between two instances when running
    with two different weights in a linear combination on the ChatBot Trace using the Qwen3-30B model.
    The reported metric is the absolute served prefill time in each 10-second window
    between the two instances (Inst.).
    }}
    \label{fig:chln-pt-imbalance}
\end{figure}

\nospacestitle{A starting case: the vLLM policy~\cite{vllm-code}. \,}
Our study starts with the default global scheduling policy adopted by
vLLM~\cite{vllm-code}---a popular open-source LLM serving engine
widely used in both industry and academia.
{\fig{fig:vllm-kvcache}} (a) shows its scheduling method,
which uses the batch size of each instance as the indicator for the routing score.
It is a variant of the classic load-balancing-centric join-the-shortest-queue (JSQ) policy
with an extension for LLM: at each instance,
the batch size includes both the requests running on the instance (R-BS)
and those queued in the instance's queue (Q-BS).

\begin{figure*}[!t]
    \centering
     \includegraphics[width=0.95\linewidth, trim=0.25cm 14.52cm 26.5cm 0.25cm, clip]{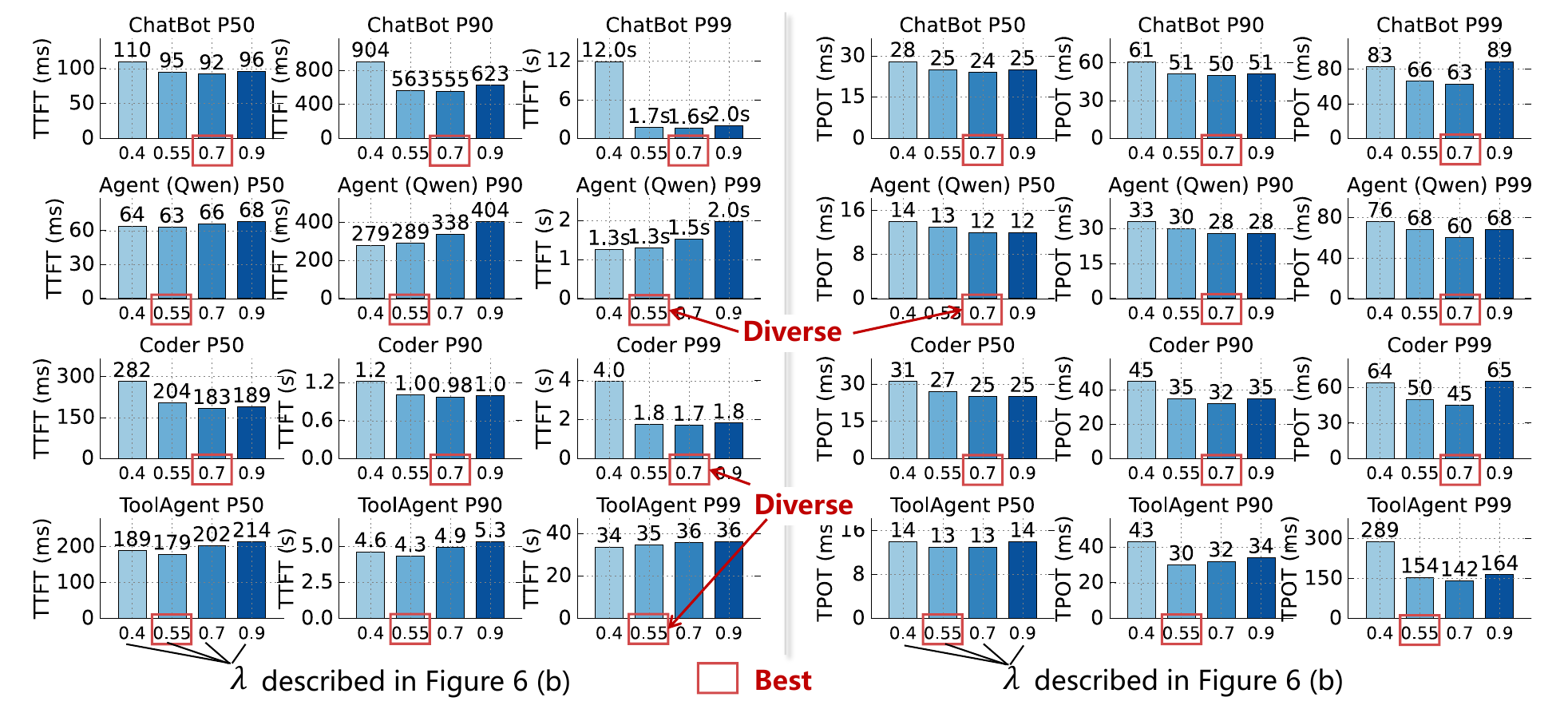} \\[-1pt]
    \begin{minipage}{1\linewidth}
    \caption{\small{
        An analysis of how the performance 
        varies with different hyperparameters on 
        four traces with a linear-combination-based method.
        The model used is Qwen3-30B. 
        }}
    \label{fig:data-linear-tune}
    \end{minipage} \\[-10pt]
\end{figure*}

\stitle{Retrofitting vLLM with {\kvcache}-awareness (\company). \,} 
While JSQ tries to balance the workload across instances,
it is unaware of the {\kvcache} state of the incoming requests.
This matters because routing requests to instances with a higher {\kvcache} hit rate
substantially reduces request latency.
{\fig{fig:vllm-kvcache}} (b) shows a simple extension of vLLM
adopted by {\company} and others~\cite{ai-Dynamo,aigw} to make request scheduling {\kvcache}-aware:
it adds a {\kvcache} indicator to the score function---an estimate
of the {\kvcache} hit ratio if the request is routed to that instance.
Here, $KV$ is the symbolic value representing the per-instance {\kvcache} hash map.
There are two points to note here:
(1) the load balance indicator (batch size) needs to be normalized to $[0,1]$
to match the scale of the hit ratio because otherwise
they cannot simply be added together;
(2) the analysis in this section assumes the linear combination coefficients ($\lambda$) are fixed,
and the next section discusses the rationale for setting them.

As shown in {\fig{fig:kvcache-aware-cdf-1}}, 
adding {\kvcache}-awareness to a load-balancing-only policy 
improves the average TTFT by 84\% and the average TPOT by 17\%, 
which is as expected because 
it increases the {\kvcache} hit ratio as profiled in {\fig{fig:kvcache-aware-hit-ratio1}}. 
Interestingly, although the improved {\kvcache} hit ratio---at first glance---is only
beneficial to the prefill, 
we found it also reduces the decode time
because it reduces the computation required for each instance, 
allowing the instance to dedicate more GPU time to the decode phase.

\subsection{{\kvcache}-awareness vs. Load balancing: The Trade-off}
\label{sec:challenge-balancing}

\noindent
Although adding {\kvcache}-awareness is intuitively beneficial (\textsection{\ref{sec:kvcache-aware}}), realizing this benefit in practice is non-trivial because the {\kvcache} objective may interfere with load balancing.
With a linear combination, the priority of each objective is controlled by the weight assigned to each indicator ($\lambda$ in {\fig{fig:vllm-kvcache}} (b)):
a larger weight on the {\kvcache} component prioritizes routing requests to instances with higher hit ratios even when other instances are much less loaded, leading to load imbalance.

{\fig{fig:data-linear-tune}} illustrates this trade-off, reporting the overall TTFT and TPOT by sweeping weights.
We can see that when increasing the weight from 0.4 to 0.9, the TTFT first gradually decreases and then increases, except for the API trace, which is less affected due to its short input length.
The increased {\kvcache} hit ratio explains this: as shown in {\fig{fig:chln-cache-hit}}, on the ChatBot trace (other traces are similar) the hit ratio rises accordingly with the increased weight. 

Despite the increased {\kvcache} hit ratio, there is a knee point in the weight (e.g., 0.7 for ChatBot), beyond which the overall latency starts to increase due to load imbalance across instances.
{\fig{fig:chln-pt-imbalance}} profiles this imbalance, plotting the prefill work assigned to two instances under different weights (0.7 vs. 0.9) over the same 400-second burst period on the ChatBot trace.
The prefill time---seconds spent on prefill within each 10-second window---measures the workload per instance: since most tokens are generated during decode, an instance dominated by prefill produces fewer tokens than others.
For each setup, we select the two instances (out of 16) with the highest standard deviation of prefill time.
We observe that under $\lambda=0.9$ the average prefill time differs significantly between the two instances (3.57s vs. 2.17s), while $\lambda=0.7$ yields similar values (3.43s vs. 3.40s).

\subsection{The Case of Linear Combination}
\label{sec:linear-combination}

\noindent
Based on the trade-offs explored previously, 
using a linear combination to achieve both {\kvcache}-awareness and load balancing leads to two issues:

\stitle{Cons \#1. Requires workload-specific hyperparameter tuning. \,}
The importance of {\kvcache} and its impact on load imbalance are workload-dependent.
For example, {\fig{fig:data-linear-tune}} presents the tuning results for different traces when running a Qwen3-30B model.
We can see that the evaluated optimal weight for each workload varies:
in ChatBot the optimal weight is 0.7, while in API it is 0.55,
even though both workloads have a similar {\kvcache} access pattern.
Note that we cannot afford to sweep all possible configurations,
as replaying each trace on the testbed consumes a substantial amount of GPU time.
As a result, the linear combination requires workload-specific hyperparameter tuning that
is non-trivial in practice due to the diversity of workloads~\cite{10.5555/3768039.3768067}.

\stitle{Cons \#2. Sub-optimal performance. \,}
During the evaluation in \textsection{\ref{sec:eval}},
we found that a statically tuned weight cannot always achieve competitive performance compared to other baselines.
We hypothesize that this is because the optimal weight may vary over time.
Intuitively, if the GPUs are idle, we need to prioritize {\kvcache} hits
with a larger {\kvcache} weight. On the other hand, if prioritizing {\kvcache}
results in an imbalance,
we need to reduce the weight to improve load balancing.
However, to the best of our knowledge,
all existing work uses a fixed tuned weight for the entire serving duration.
While it is possible to design strategies to adaptively tune the weight over time,
doing so would add system complexity.

\begin{figure*}[!t]
    \centering
     \includegraphics[width=0.99\linewidth]{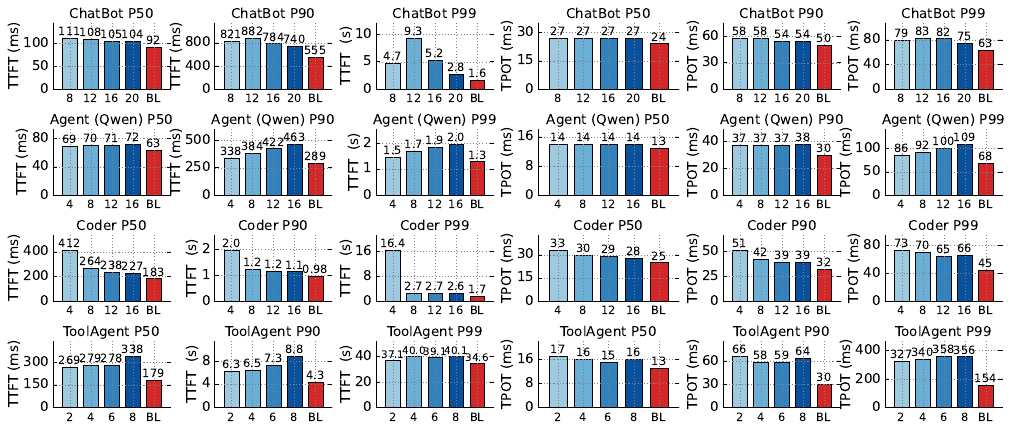} \\[2pt]
    \begin{minipage}{1\linewidth}
    \caption{\small{
An analysis of how performance
        varies with different hyperparameters on
        four traces using a filter-based combination method.
        \textbf{BL} denotes the \uline{l}inear-combination-based method for comparison, tuned with the \uline{b}est
        hyperparameter.
        The numbers (2,4,6,8) are the range described in {\fig{fig:aibrix}}.
        The model used is Qwen3-30B. 
        }}
    \label{fig:data-filter-tune}
    \end{minipage} \\[-10pt]
\end{figure*}

\subsection{The Case of Filter-based Combination}
\label{sec:filter-combination}

\begin{figure}[!t]
    \centering
     \includegraphics[width=0.9\linewidth, trim=0.25cm 18.9cm 37.65cm 0.25cm, clip]{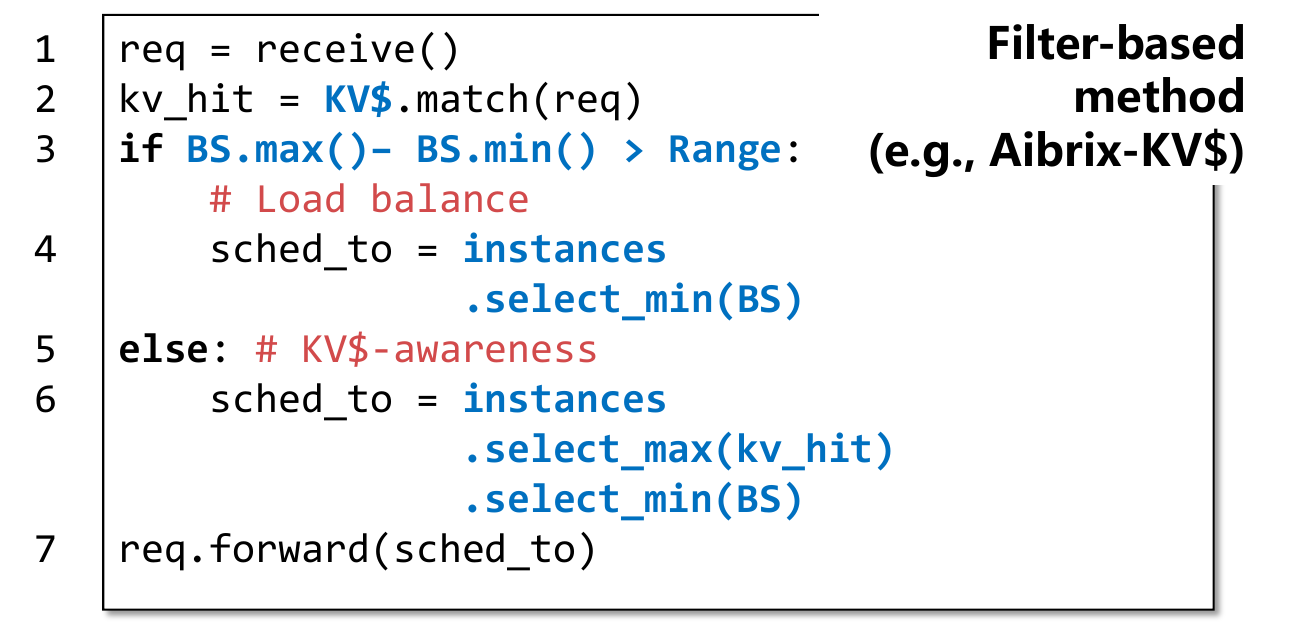} \\[5pt]
    \begin{minipage}{1\linewidth}
    \caption{\small{
        The pseudocode of filter-based combination of {\kvcache}-awareness and load balancing in LLM scheduling,
        simplified from \texttt{prefix-cache} policy of AIBrix~\cite{aibrix}.
        }}
    \label{fig:aibrix}
    \end{minipage} \\[-10pt]
\end{figure}

\nospacestitle{Methodology. \,}
{\fig{fig:aibrix}} shows how a typical filter-based combination works for
integrating {\kvcache}-awareness and load balancing
in LLM scheduling systems such as AIBrix~\cite{aibrix}.
First, the router checks whether the current cluster has an imbalanced load---i.e.,
the range between the maximum and minimum batch sizes across instances (\(\texttt{BS.max() - BS.min()}\)) exceeds a threshold (line 3).
If so, the router abandons {\kvcache}-awareness and simply routes the request to the instance
with the smallest batch size for load balancing (lines 4--5).
Otherwise, the router uses the {\kvcache} hit ratio as an indicator to route requests to instances (lines 6--9) for {\kvcache}-awareness.

\stitle{Cons \#1. Still requires workload-specific hyperparameter tuning. \,}
Similar to linear combination, filter-based methods also require
hyperparameter tuning because the threshold for determining load imbalance (\texttt{Range})
is workload-dependent.
As shown in {\fig{fig:data-filter-tune}},
the optimal threshold of a typical filter-based method~\cite{aibrix} varies across workloads:
for example, in Coder, increasing the threshold from 4 to 16 improves
the P50 TTFT and TPOT by 44\% and 15\%, respectively. 
On the other hand, for the API trace, 16 is a better choice than 4.

\stitle{Cons \#2. Sub-optimal performance. \,}
Besides hyperparameter tuning, filter-based combination is slower
than linear combination with properly tuned weights,
as shown in {\fig{fig:data-filter-tune}}, 
because it biases towards load balancing and may forgo the benefits of {\kvcache}-awareness. 
Specifically, when a load imbalance is detected, it completely ignores {\kvcache}-awareness,
even though routing requests to instances with a higher {\kvcache} hit ratio could still be beneficial
if the amount of reduced computation is significant (as it helps reduce the load).
With linear combination, this is possible as long as the weight
assigned to the {\kvcache} hit ratio is not too small. 
This is not possible in filter-based methods. 

\subsection{The Case of Simulation-based Combination}
\label{sec:prediction-combination}
\begin{figure}[!t]
    \centering
     \includegraphics[width=0.96\linewidth, trim=0.25cm 25cm 38cm 0cm, clip]{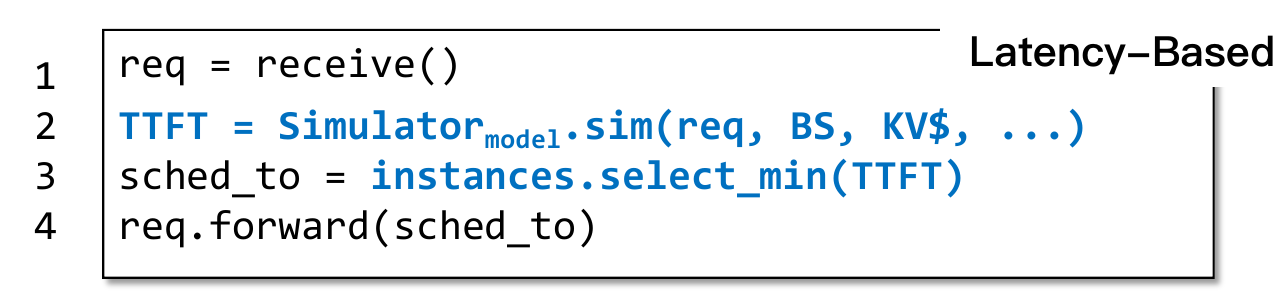} \\[6pt]
    \begin{minipage}{1\linewidth}
    \caption{\small{
        The pseudocode of simulation-based method for combining {\kvcache}-awareness and load balancing in LLM scheduling. 
        }}
    \label{fig:latency-based-scheduling}
    \end{minipage} \\[-10pt]
\end{figure}

\begin{figure*}[!t]
    \centering
    \includegraphics[width=0.85\linewidth]{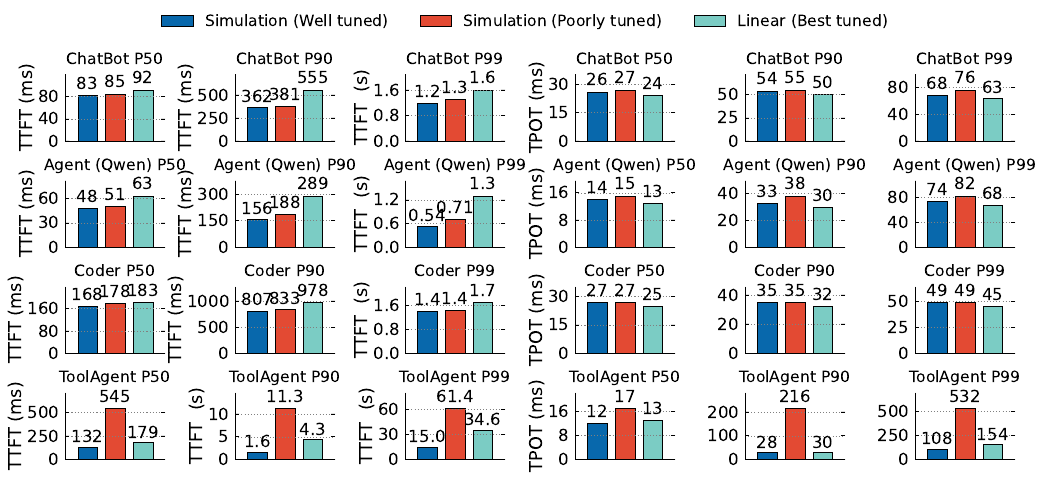} \\[-5pt] 
    \begin{minipage}{1\linewidth}
    \caption{\small{
    An analysis of how performance varies with different simulator accuracy across
    four traces using a Qwen3-30B model.
        }}
    \label{fig:wrong-predictor-e2e}
    \end{minipage} \\[-15pt]
\end{figure*}

\begin{figure}[t]
    \centering
    \includegraphics[width=0.85\linewidth]{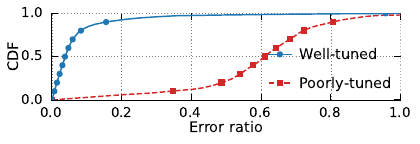} \\[-5pt]
    \caption{\small{   
    A comparison of a well-tuned simulator
    vs. a non-tuned one on ChatBot Trace with Qwen3-30B model. 
    }}
    \label{fig:predictor-accuracy-cdf}
\end{figure}

\nospacestitle{Methodology. \,}
Finally, simulation-based methods~\cite{305212, llm-d, DBLP:journals/corr/abs-2507-17769, DBLP:journals/corr/abs-2504-08784}
use the simulated serving time of routing a request to a specific instance as the routing score.
This is based on the observation that the execution of LLM requests is relatively deterministic---each step is a single forward pass---so it can be simulated accurately.
{\fig{fig:latency-based-scheduling}} shows the pseudocode of a typical simulation-based method~\cite{llm-d}:
after receiving a request,
the router first estimates the TTFT of routing the request to each instance via a simulator (e.g., VIDUR~\cite{DBLP:conf/mlsys/AgrawalKMPKGRT24}) (line 2),
and then routes the request to the instance with the lowest estimated TTFT (lines 3--4).
Note that simulation-based methods typically estimate the TTFT instead of the end-to-end latency
because the end-to-end latency depends on the number of output tokens, which is unpredictable.

Simulation-based methods can be viewed as a higher-order combination
of indicators that achieve {\kvcache}-awareness and load balancing.
This is because the simulator must use the {\kvcache} state and batch size of each instance
as input features to accurately simulate the TTFT.
Here, we chose a simulation-based implementation 
similar to llm-d~\cite{llm-d}
 but with a retrofitted simulator from VIDUR~\cite{DBLP:conf/mlsys/AgrawalKMPKGRT24}---the 
state-of-the-art LLM instance simulator. 
Our retrofitting consists of two parts:
(1) we extend the simulator to consider {\kvcache}-aware execution by modeling the prefill phase with cache hits,
and (2) we re-implement it in Rust and enable parallel simulation to scale the online simulation to multiple instances. 
Without {\kvcache}-awareness, the simulation-based method does not outperform its counterparts in our setting,
similar to the observations we made in \textsection{\ref{sec:kvcache-aware}}.
Without Rust, the original Python-based implementation has considerable scheduling latency
and would incur substantial online scheduling overhead.
We also have several optimizations to scale the simulator to hundreds of instances.
We will leave the details to another paper as it is not the focus of this work. 

We studied a state-of-the-art method and found that it
outperforms the linear-combination-based method in certain traces,
as also shown in {\fig{fig:latency-based-scheduling}}.
Despite the improved performance, simulation-based methods still have two issues due to their complexity:

\stitle{Cons \#1. Development complexity due to per-model implementation and per-hardware tuning.  \,}
The performance of simulation-based methods depends on the accuracy of the simulator,
which is non-trivial to achieve in practice because
we need to consider both the model architecture and the hardware characteristics.
To quantify the impact of simulator accuracy on scheduling performance,
{\fig{fig:wrong-predictor-e2e}} presents the performance of
using well-tuned vs. non-tuned simulators on four traces when serving a Qwen3-30B model.
The poorly tuned simulator is one originally used for another model---Qwen2-7B,
while a well-tuned simulator is the one implemented specifically for Qwen3-30B.
{\fig{fig:predictor-accuracy-cdf}} measures the TTFT deviation when using
two simulators to serve a Qwen3-30B model compared with using vLLM.
We can see that a well-tuned simulator achieves much higher accuracy than an untuned one.
With a more accurate simulator, the TTFT and TPOT tail latency improve
by 75.6\% and 79.7\%, respectively.

While it is possible to develop per-model simulators for high scheduling performance,
doing so incurs non-trivial development complexity
because the simulations 
highly depend on the model architecture---which is evolving rapidly
with new modules (e.g., linear attention~\cite{qwen3-next} and Engram~\cite{cheng2026conditionalmemoryscalablelookup}).
Meanwhile, for the same model, the hyperparameters of the simulator need to be tuned according to the hardware characteristics,
leading to further engineering efforts.

\stitle{Cons \#2. Still sub-optimal performance. \,}
Simulation-based methods still suffer from sub-optimal performance,
especially for TPOT.
As shown in {\fig{fig:wrong-predictor-e2e}},
on the ToolAgent trace, the TPOT tail latency is 71.1\% slower than that of the best linear-combination-based method.
We hypothesize that this is because mispredictions of the simulator lead to load imbalance.
For example, in {\fig{fig:predictor-accuracy-cdf}}, we can see
that even with a well-tuned simulator, there are still about 10\% of requests with more than 20\% prediction error.
Such errors mainly come from two sources: request reordering at the vLLM API server,
and inaccuracies in latency prediction.

%% file: method-v1.tex
\section{Simple Multiplication May Be All You Need}
\label{sec:method} 

\begin{figure}[!t]
        \begin{minipage}{1\linewidth}
        \centering
        \includegraphics[width=0.92\linewidth, trim=0.25cm 13.5cm 38.4cm 0.25cm, clip]{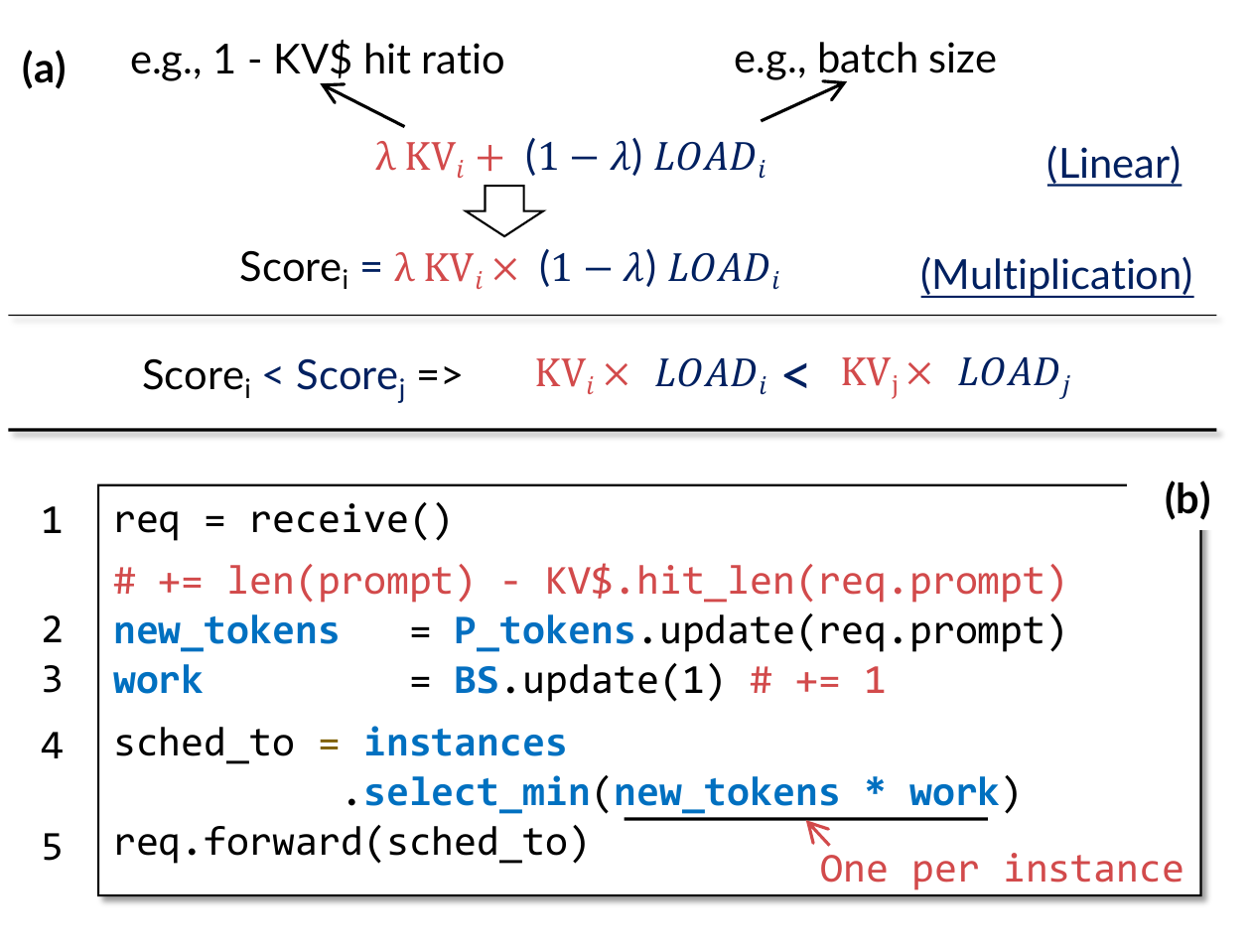}
        \end{minipage} \\[-1pt]
        \begin{minipage}{1\linewidth}
        \caption{\small{%
    (a) An illustration of how multiplication combines two indicators without the hyperparameters needed in linear combination, 
    and (b) the pseudocode of our scheduling method.
        }}  
        \label{fig:method}
        \end{minipage} \\[-10pt]
        \end{figure}

\nospacestitle{The methodology ({\fig{fig:method}}). \,}
Our method is simple: we only need to multiply two carefully chosen indicators to compute the scheduling score,
one for {\kvcache}-awareness and the other for load balancing,
and then route the request to the instance with the minimal score.
The basic idea is based on the observation that,
if a linear combination of two indicators works, then multiplication can also work, with the benefit of avoiding hyperparameter tuning,
as shown in (a). 
The two indicators---\textbf{P-token} (the number of new prefill tokens if the request is routed to an instance, considering {\kvcache} hits) and \textbf{BS} (the batch size of the instance)---are chosen based on our analysis in \textsection{\ref{sec:indicators}}.

To see why routing to the instance with the minimal (P-token $\times$ BS) score considers both objectives well,
consider two instances $i$ and $j$:
if routing the request to instance $i$ results in more {\kvcache} hits than routing it to instance $j$,
then instance $i$ will have a lower \textbf{P-token} value unless there are many queued prefill requests in instance $i$
 (indicating work imbalance).
Meanwhile, the \textbf{BS} captures the decode workload of each instance,
so if instance $i$ has a significantly larger batch size than instance $j$,
the multiplication will likely favor instance $j$ for load balancing.

\subsection{The Choice of the Indicators}
\label{sec:indicators}

\begin{figure}[!t]
        \begin{minipage}{1\linewidth}
        \centering
        \includegraphics[width=0.95\linewidth, trim=0.25cm 13.1cm 40.5cm 0.25cm, clip]{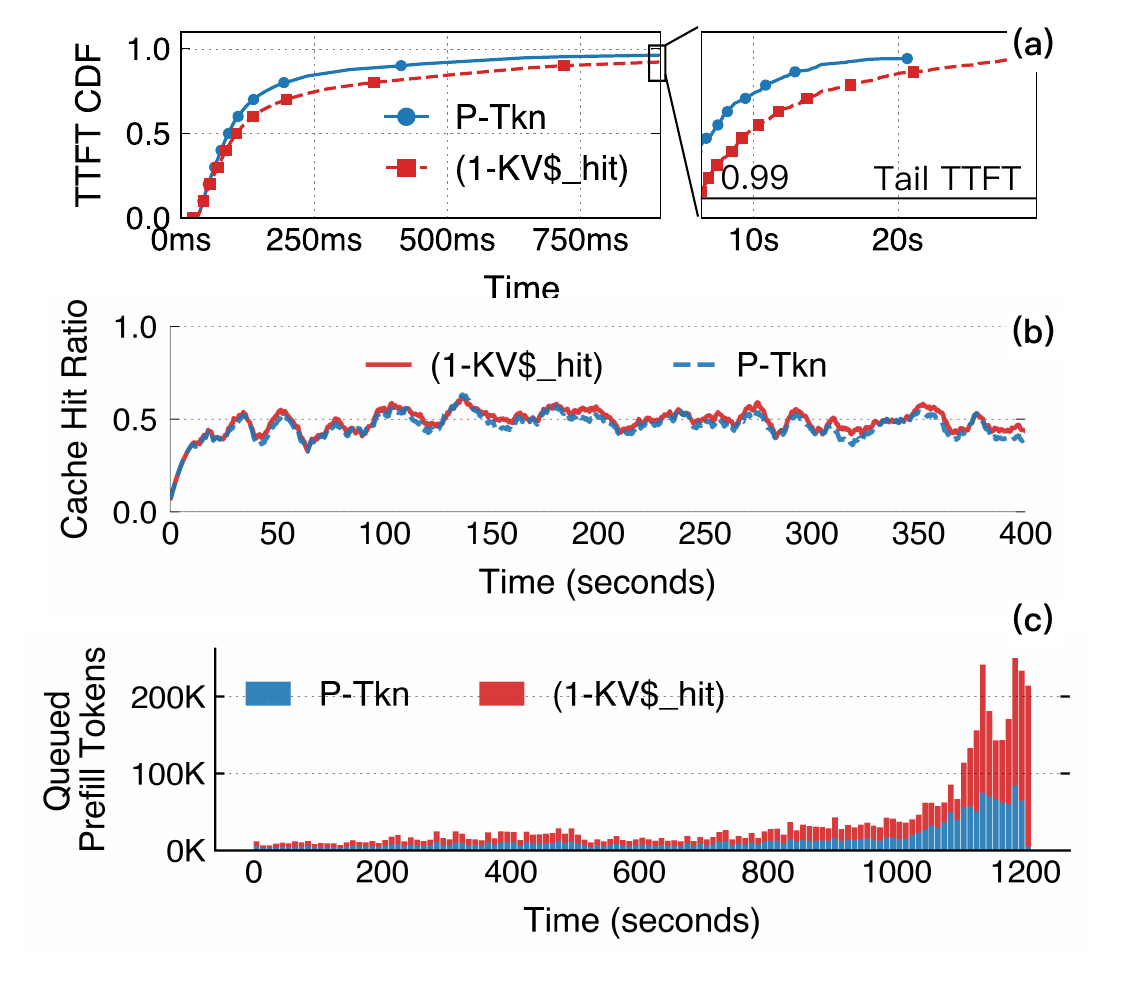}
        \end{minipage} \\[-1pt]
        \begin{minipage}{1\linewidth}
        \caption{\small{%
    (a) A comparison of using new prefill tokens (\textbf{P-token}, shown as ``P-Tkn'' in the figure) vs. 1-{\kvcache} hit ratio (1-KV$_{hit}$)
    as the {\kvcache}-awareness indicator ($A$) in $A \times BS$ scheduling,
    (b) the {\kvcache} hit ratio analysis, and
    (c) the queued prefill tokens analysis.
    The analysis is done on the Qwen3-30B model with the ChatBot(Qwen) trace.
    Note that the bottom graph is not a stacked graph; rather, the bars overlap each other.
        }}  
        \label{fig:whyA}
        \end{minipage} \\[-15pt]
        \end{figure}

\nospacestitle{{\kvcache}-awareness indicator (P-token). \,}
Besides \textbf{P-token}, another natural choice for the {\kvcache}-awareness indicator is
the \textbf{1-{\kvcache} hit ratio}, i.e., the {\kvcache} hit ratio if the request is routed to an instance.
It is adopted by works such as Preble~\cite{DBLP:conf/iclr/SrivatsaHAL025} and AIGW~\cite{aigw}.
Note that we subtract from one because a higher {\kvcache} hit ratio should yield a lower score
to align with the new prefill tokens indicator.
We do not consider TTFT because it requires a simulator,
which is not always applicable.

Our empirical analysis shows that using \textbf{P-token} yields better performance than using \textbf{1-{\kvcache} hit ratio}.
{\fig{fig:whyA}} (a) shows that using \textbf{P-token} results in a 14.4\% lower
P50 TTFT and a 42.8\% lower P95 TTFT compared to using \textbf{1-{\kvcache} hit ratio}.
For a fair comparison, we fix the load-balancing indicator to \textbf{BS} in both cases.
Due to space limitations, we only report the results for one workload here;
however, the trend is consistent across all evaluated workloads.

To understand the cause of this difference,
{\fig{fig:whyA}} further breaks down the {\kvcache} hit ratios of different methods in (b)
and the load balancing status in (c).
We can see that the two methods achieve similar {\kvcache} hit ratios, so
they are equally {\kvcache}-aware.
The key difference is that using \textbf{P-token} achieves better load balancing,
as it additionally considers the queued prefill tokens in each instance.
As a result, when making scheduling decisions, the router bypasses instances
with many queued prefill requests, even if they have a high {\kvcache} hit ratio.

\begin{figure}[!t]
        \hspace{-2mm}
        \begin{minipage}{1\linewidth}
        \centering
        \includegraphics[width=0.99\linewidth, trim=0.25cm 18.5cm 39.5cm 0.25cm, clip]{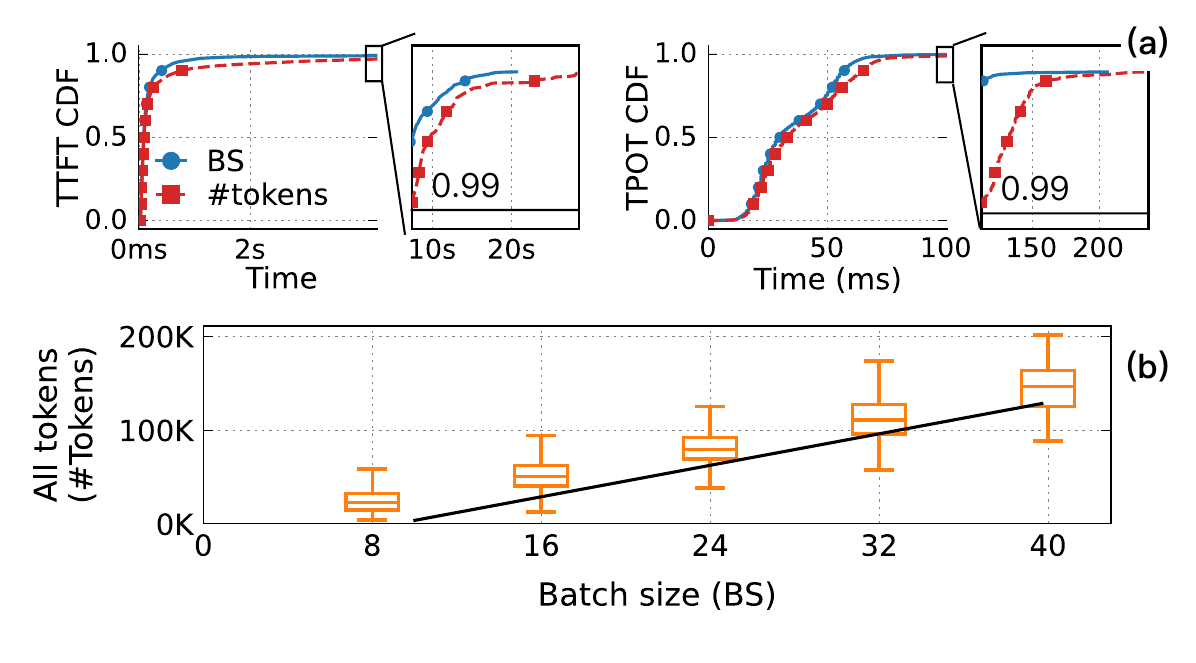}
        \end{minipage} \\[-2pt]
        \begin{minipage}{1\linewidth}
        \caption{\small{%
    (a) A comparison of using batch size (BS) vs. total tokens (\#Tokens)
    as the load-balancing indicator ($B$) in P-token $\times B$ scheduling.
    (b) The profiled relationship between batch size and total tokens.
    The analysis is done on the Qwen3-30B model with the ChatBot(Qwen) trace.
        }}  
        \label{fig:whyB}
        \end{minipage} \\[-10pt]
        \end{figure}

\stitle{Load-balancing indicator (BS). \,}
Besides \textbf{BS}, another common choice is the
total tokens (\textbf{\#Tokens}) on each instance, as adopted by works such as ai-Dynamo~\cite{ai-Dynamo} and AIGW~\cite{aigw}.
The rationale is that the total computation cost of a request is proportional
to the number of tokens in its context.
However, we found that using \textbf{BS} yields better performance,
as shown in {\fig{fig:whyB}} (a), for two reasons.
First, workloads can be categorized into prefill and decode loads,
where the former is considered in our \textbf{P-token} indicator.
Thus, we only need an indicator for the decode workload, which is exactly what \textbf{BS} captures (note that we
have also tried using decode batch size, and the results are similar
since the \textbf{BS} is dominated by the decode requests).
Moreover, \textbf{BS} is a better indicator for the (decode) work assigned to an instance because the decode time
is more stable across batch sizes.
Specifically, a larger batch size leads to a longer decode time,
but more decode tokens do not necessarily mean decode is slower when the batch is small,
thanks to the {\kvcache}~\cite{DBLP:conf/osdi/ZhongLCHZL0024}.
This is illustrated in {\fig{fig:whyB}} (b),
where we profile the relationship between batch size and total tokens
while serving a ChatBot(Qwen) workload with Qwen3-30B.

\subsection{Benign and Failure Cases Analysis of the Multiplicative Scheduling Score}
\label{sec:m-analyze}

\nospacestitle{Overview. \,}
At a high level, as long as there is no load imbalance,
i.e., one instance is overloaded while others are idle,
routing requests to instances with high {\kvcache} hit rates (i.e., low \textbf{P-token})
is always beneficial.
Thus, the multiplication fails when a set of instances is about to be overloaded
but their increase in \textbf{BS} cannot offset the decrease in \textbf{P-token} from high {\kvcache} hit rates,
so requests keep being routed there and cause the imbalance.
This can happen under extreme {\kvcache} skewness:
when some prefixes are repeatedly accessed but only cached 
on a small set of instances---which we term {\kvcache} \emph{hotspots}.
We found that such hotspots are rare in practice---at least not
present in any of our evaluated traces.
More importantly, as their patterns are clear, we can design a detector to catch them beforehand.

To analyze the failure cases of our multiplication method,
we derive the condition under which {\kvcache} hotspots occur and
the increased batch size cannot offset the high {\kvcache} hit rate given a workload pattern.
For each workload, we first group requests by their {\kvcache} prefixes
and partition instances according to their {\kvcache} ownership.
Next, for each request class, we analyze the
relationship between the workload pattern, the P-token indicator given the {\kvcache} hit rate,
and the batch size indicator.

\stitle{Prelude: request grouping and instance partitioning. \,}
First, we partition all requests into a set of classes
\(\mathcal{C}\), where each \(c\!\in\!\mathcal{C}\) corresponds to a
group of requests that share the same {\kvcache} prefix.
In practice, a class roughly matches an application or a user: their
requests share the same system prompt and often a similar conversation
history~\cite{10.5555/3768039.3768067}.
For a fixed class \(c\), we consider an accumulation time window
\((t, t+\textit{window})\) and denote by \(x\) the fraction of all
requests arriving at the cluster that belong to class \(c\) within this window; the
remaining fraction is \(\overline{x} = 1-x\).

For each class \(c\), we partition the cluster into two sets of instances: 
\(M\) and \(\overline{M}\).
\(M\) contains instances whose cache currently holds the prefix of
class \(c\), i.e., with {\kvcache} hits, while \(\overline{M}\) contains all other instances.

\stitle{Impact of a request class on batch size. \,}
Let \(\text{QPS}\) be the total query rate of the cluster, 
and let \(t\) be the arrival time of the suspected class-\(c\) requests that may lead to imbalance.
We denote the average batch size per instance 
in a balanced state prior to \(t\) as \(BS_0\).
We also denote the expected batch sizes of instances in \(M\) and \(\overline{M}\) by 
\(BS_{t}\) and \(\overline{BS}_{t}\) during the period \((t, t+\textit{window})\). 
We assume the extreme case in which all class-\(c\) requests are routed to \(M\) due to high {\kvcache} hits,
because routing any class-\(c\) request to an instance in \(\overline{M}\) would cache the prefix there, 
expand \(M\), and thereby dissipate the hotspot.
Under this worst-case assumption, we establish the following
approximate expression for the ratio between a potentially overloaded
instance's batch size and a non-overloaded one's:

\begin{equation}
\frac{BS_{t}}{\overline{BS}_{t}}
  = \frac{
      BS_0 + \left( x \cdot \text{QPS}\right) / |M| \cdot t
    }{
      BS_0 + \left( \overline{x} \cdot \text{QPS}\right) / |\overline{M}| \cdot t
    }.
\label{eq:bs-ratio-exact}
\end{equation}
The main term
\(\left( x \cdot \text{QPS}\right) / |M| \cdot t\) corresponds to the number of 
requests of class \(c\) routed to the instances that cache its prefix,
while \(\left( \overline{x} \cdot \text{QPS}\right) / |\overline{M}| \cdot t\) is the
number of all other requests routed to the remaining instances.

\stitle{Analysis: when multiplication suffices, and is this common? \,}
As long as the batch size of the suspected hotspot instances is not
larger than that of the others, it is beneficial to route requests to these
instances, since doing so can exploit the {\kvcache} hits without
incurring load imbalance.
This is precisely what our multiplicative method is designed for.
The remaining question is whether such cases are common in practice.
Fortunately, Equation~\ref{eq:bs-ratio-exact} reveals that the
batch-size skewness is governed by two ratios: the class popularity
\(x/\overline{x}\) and the cache coverage \(|M|/|\overline{M}|\).
We can therefore empirically analyze the prevalence of such cases by
tracking these two ratios at runtime, sampling the request classes
with the highest {\kvcache} hits within each one-minute window.

\begin{figure}[!t]
          \begin{minipage}{1\linewidth}
          \centering
          \includegraphics[width=0.92\linewidth]{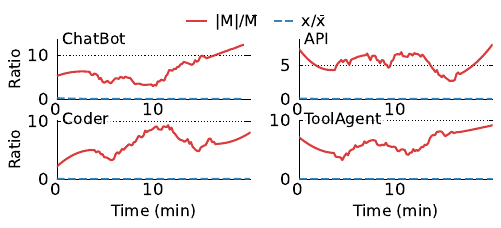}
          \end{minipage} \\[-0pt]
          \begin{minipage}{1\linewidth}
          \caption{\small{%
Empirical observations of the factors in the multiplicative score
across different traces.
If $x/\overline{x} \leq |M|/|\overline{M}|$, no {\kvcache} hotspot can
cause load imbalance, so our multiplicative method can effectively
balance the load with {\kvcache}-awareness.
          }}
          \label{fig:detector}
          \end{minipage} \\[-10pt]
          \end{figure}

{\fig{fig:detector}} shows that for all traces (representative of LLM
serving and studied in this work), our multiplication remains in its
applicable regime, because the expected batch size of the suspected
hotspot instances is not larger than that of the others.
Concretely, every sampled class satisfies Equation~\ref{eq:consensus-1}:

\begin{equation}
\frac{x}{\overline{x}}
\leq
\frac{|M|}{|\overline{M}|}
\label{eq:consensus-1}
\end{equation}

\noindent
which says the class's relative popularity (\(x/\overline{x}\)) never
exceeds the relative cache coverage it enjoys
(\(|M|/|\overline{M}|\)); i.e., the prefix is cached on enough
instances to absorb its share of the arrivals.
This implies that even when every class-\(c\) request lands on \(M\), no hit
instance accumulates a larger batch size than a non-hit one, 
i.e., 
transforming and then substituting Equation~\ref{eq:consensus-1} into
Equation~\ref{eq:bs-ratio-exact}, we get
\(BS_t \leq \overline{BS}_t\).

\begin{figure}[!t]
          \centering
          \begin{minipage}{1\linewidth}
          \centering
          \includegraphics[width=0.92\linewidth]{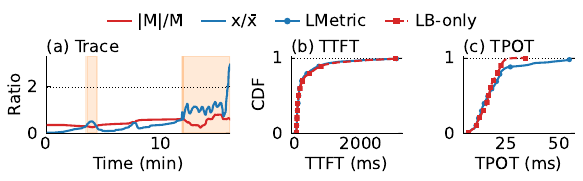}
          \end{minipage} \\[2pt]
          \begin{minipage}{1\linewidth}
          \caption{\small{%
Adversarial case study: (a) $x/\overline{x} > |M|/|\overline{M}|$ implies a hotspot that could cause imbalance
under {\sys}, (b--c) performance comparison with a load-balance-only solution. 
          }}
          \label{fig:adversarial}
          \end{minipage} \\[-0pt]
          \end{figure}

\stitle{Analysis: when multiplication fails. \,}
Based on the previous analysis, if Equation~\ref{eq:consensus-1} does
not hold, i.e., \(x/\overline{x} > |M|/|\overline{M}|\), then a
{\kvcache} hotspot can cause load imbalance that breaks our
multiplicative method.
Though we have not observed such cases in any of our evaluated traces,
we used the pattern of {\kvcache} hotspots to examine a subset of
the production workloads in {\company} and found one, 
as shown in the orange-shaded windows in {\fig{fig:adversarial}} (a).
The cluster serves a thinking workload
with bursts of long requests sharing a common prefix,
visible around minute~11 of the run.
In this case, the batch indicator cannot outperform the {\kvcache} indicator,
and thus the multiplicative score continuously routes requests to a few instances with high {\kvcache} hits,
leading to load imbalance and performance degradation.
{\fig{fig:adversarial}} (b--c) shows that {\sys} cannot outperform a load-balancing-only solution (i.e., vLLM) during
this period.

\stitle{Retrofit: a two-phase detector for {\kvcache} hotspots. \,}
The boundary condition in Equation~\ref{eq:consensus-1} yields a
clear failure-case detector based purely on the workload pattern
and {\kvcache} states, which we run alongside each scheduling
decision to catch potential hotspots.
For each request class, we monitor the two ratios \(x/\overline{x}\)
and \(|M|/|\overline{M}|\) in real time; when
Equation~\ref{eq:consensus-1} is violated, we raise an alarm and
intervene by filtering out the suspected instances (\(M\)) from the
routing targets (e.g., the orange-shaded windows in \fig{fig:adversarial}~(a)).
To bound the monitoring overhead, we only
track requests with the highest {\kvcache} hit rates.

However, Equation~\ref{eq:consensus-1} is only a necessary (not
sufficient) condition for a hotspot, because it was derived under the
worst-case assumption that all class-\(c\) requests are routed to
\(M\).
In reality, even when Equation~\ref{eq:consensus-1} is violated, it
may still be beneficial to route requests to the suspected instances
with high {\kvcache} hits, as long as we do not route too many. 
We therefore augment the detector with a second phase: after the
first phase raises an alarm, we track each subsequent class request
and filter out \(M\) from scheduling only when \(2|M|\) consecutive requests would
receive a smaller multiplicative score on a hotspot instance
\(m \in M\) than on a non-hotspot one \(m' \in \overline{M}\):
i.e., \(\text{P-token}_m \times BS_m \leq
\text{P-token}_{m'} \times BS_{m'}\).

%% file: eval.tex
\section{End-to-end Evaluation}
\label{sec:eval}

\subsection{Comparison with Production Schedulers}
\label{sec:eval-industrial}

\begin{figure}[!t]
    \centering
    \begin{subfigure}{\linewidth}
        \centering
        \includegraphics[width=0.98\linewidth, trim=0.25cm 25.2cm 43.75cm 0.25cm, clip]{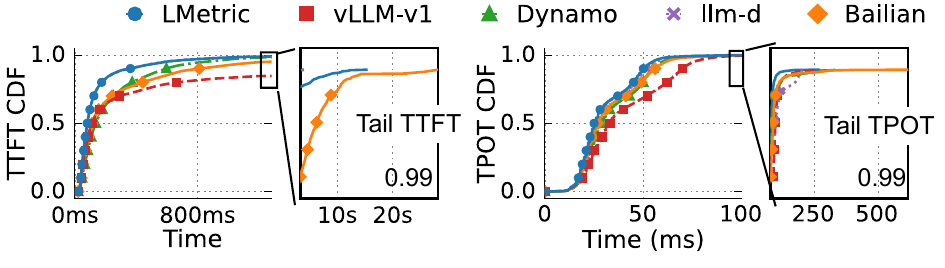}
        \vspace{-8pt}
        \caption{\small{Qwen3-30B on ChatBot (Qwen) workload.}}
        \label{fig:eval-qwen3-to-c}
    \end{subfigure} \\[6pt]
    \begin{subfigure}{\linewidth}
        \centering
        \includegraphics[width=0.98\linewidth, trim=0.25cm 25.2cm 43.75cm 0.25cm, clip]{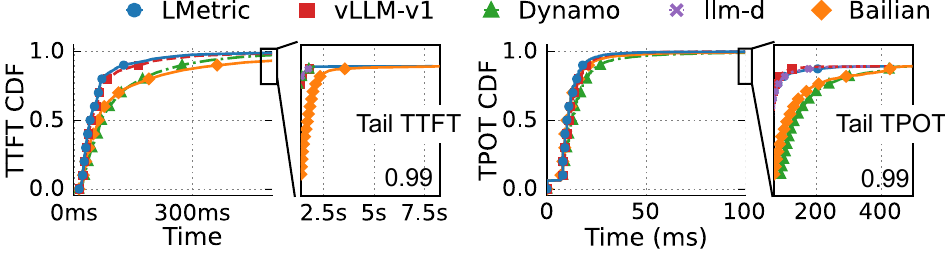}
        \vspace{-8pt}
        \caption{\small{Qwen2-7B on Agent (Qwen) workload.}}
        \label{fig:eval-qwen2-to-b}
    \end{subfigure} \\[6pt]
    \begin{subfigure}{\linewidth}
        \centering
        \includegraphics[width=0.98\linewidth, trim=0.25cm 25.2cm 43.75cm 0.25cm, clip]{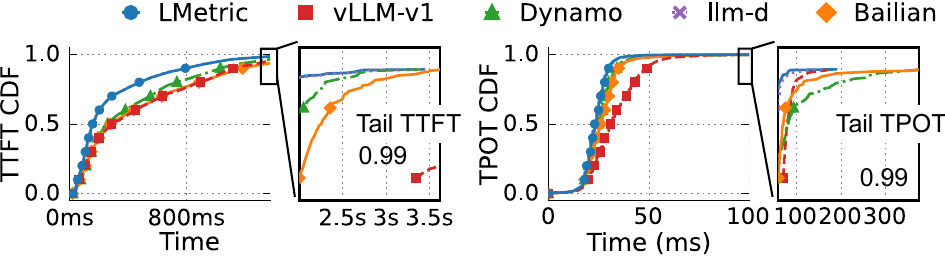}
        \vspace{-8pt}
        \caption{\small{Qwen3-30B on Coder workload.}}
        \label{fig:eval-qwen3-coder}
    \end{subfigure} \\[6pt]
    \begin{subfigure}{\linewidth}
        \centering
        \includegraphics[width=0.98\linewidth, trim=0.25cm 25.2cm 43.75cm 0.25cm, clip]{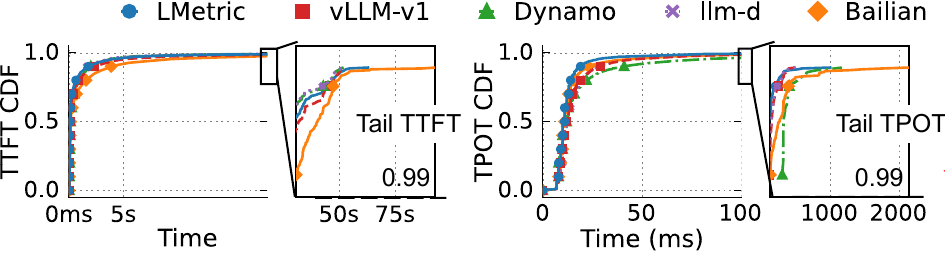}
        \vspace{-8pt}
        \caption{\small{Qwen3-30B on Agent (Kimi) workload.}}
        \label{fig:eval-qwen2-mooncake-tool}
    \end{subfigure}
    \vspace{-6pt}
    \caption{\small{End-to-end TTFT and TPOT CDFs of {\sys} and baselines on four workloads.}}
    \label{fig:eval-cdf}
\end{figure}

\begin{figure*}[!t]
    \centering
    \includegraphics[width=0.93\linewidth]{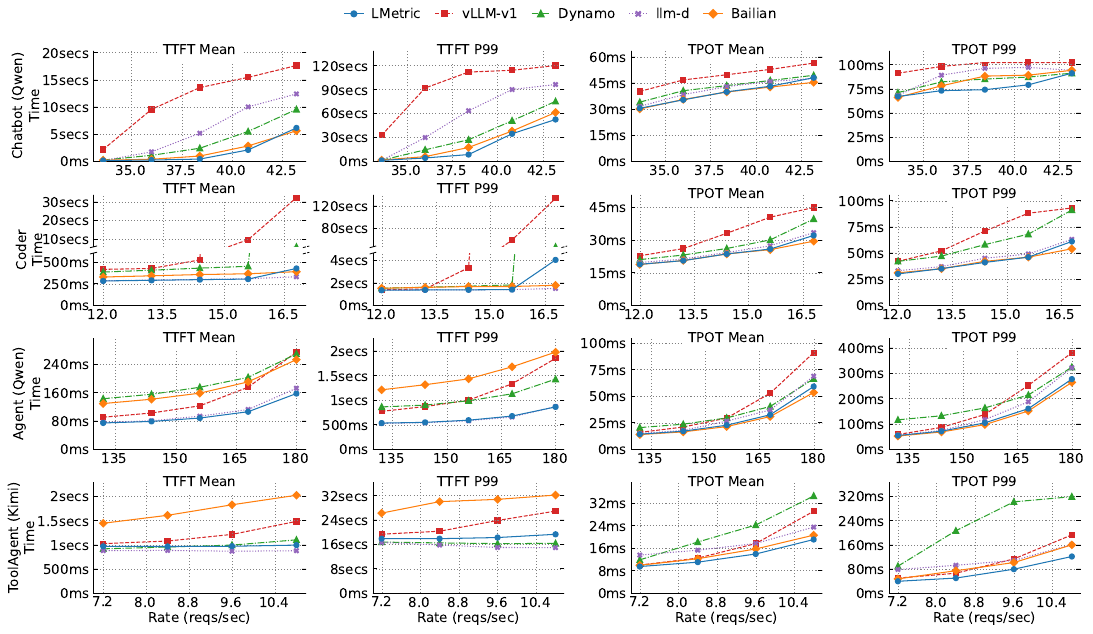} \\[2pt]
    \begin{minipage}{1\linewidth}
    \caption{\small{
    End-to-end performance under different request rates. 
    Except for the second row, which uses a Qwen2-7B model, all other
    rows use a Qwen3-30B model.
        }} 
    \label{fig:scaling-test}
    \end{minipage} \\[-15pt]
\end{figure*}

\nospacestitle{Baselines. \,}
We compare {\sys} against popular production schedulers and the current scheduler used in {\company}.
As certain baselines show low router throughput in their default implementations,
e.g., vLLM exhibits low processing throughput with its Python-based router~\cite{vllm-bug},
we re-implement their policies within our highly optimized Rust router framework described in
\textsection{\ref{sec:factory}}.
For an apples-to-apples comparison,
we evaluate all policies under our framework,
and we have carefully verified that our re-implementations are no slower than their original implementations.
The detailed descriptions of the baselines are as follows:
\\[-15pt]
\begin{itemize}[leftmargin=*,leftmargin=10pt,itemindent=0pt]
\item \textbf{\company} is the production scheduler used in {\company}'s LLM serving system. 
    It adopts a linear-combination-based approach similar to the one shown in {\fig{fig:vllm-kvcache}} (b).
    We have carefully tuned its hyperparameters for each workload to achieve the best performance.
     \\[-15pt]

\item \textbf{vLLM~\cite{vllm-code}} is a widely used LLM serving system that 
    adopts a load-balancing-only design (see {\fig{fig:vllm-kvcache}} (a)).
    
\item \textbf{Dynamo~\cite{ai-Dynamo}} is a popular serving framework
    released by NVIDIA. It also adopts a linear-combination-based approach but with a different
    choice of indicators than {\company}'s.
    The two indicators chosen are the number of prefill tokens (the same as our P-token) for {\kvcache}-awareness
    and the total tokens in the instance for load balancing.
    The router routes requests to the instance with the minimal regulated and weighted sum of the two indicators.
    Similar to {\company}, we also tune its hyperparameters for each workload to achieve the best performance. \\[-15pt]

    \item \textbf{llm-d~\cite{llm-d}} adopts a latency-based scheduling policy:
    it estimates the TTFT and routes requests to the instance with the lowest TTFT
    using the simulation-based approach described in {\textsection{\ref{sec:prediction-combination}}}. \\[-15pt]
\end{itemize}

\begin{figure}[!t]
    \centering
    \includegraphics[width=0.93\linewidth]{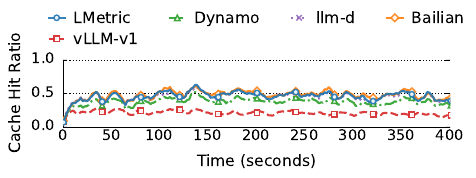} \\[-0pt]
    \begin{minipage}{1\linewidth}
    \caption{\small{   
        {\kvcache} hit ratio comparison of policies for the Qwen3-30B model on the ChatBot (Qwen) workload.
    }}
    \label{fig:eval-e2e-kvcache}
    \end{minipage} \\[-10pt]
\end{figure}

\begin{figure}[!t]
    \centering
    \includegraphics[width=0.91\linewidth]{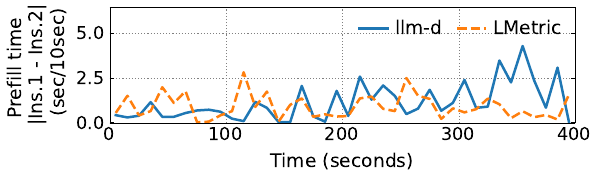} \\[0pt]
    \begin{minipage}{1\linewidth}
    \caption{\small{   
        A profile of the workload imbalance between two instances under {\sys} and llm-d
        while serving a ChatBot (Qwen) workload with the Qwen3-30B model.
        The reported metric is the absolute served prefill time in each 10-second window
        between the two instances (Inst.).
    }}
    \label{fig:eval-e2e-imbalance}
    \end{minipage} \\[-5pt]
\end{figure}

\stitle{Overall performance. \,}
{\fig{fig:eval-cdf}} shows the end-to-end
TTFT and TPOT CDFs of {\sys} and the baselines on four traces.
All experiments are conducted on a 16-GPU testbed with 16 instances,
and each trace is scaled to half of the maximum load that the testbed can handle.
Due to space constraints, we report results for a representative subset: 
the Qwen3 MoE model on ChatBot, Coder, and Agent workloads, 
and the Qwen2 model on the Agent (Qwen) workload.
We observe consistent performance trends across all model--trace combinations. 

{\sys} outperforms all baselines across all traces.
On the ChatBot workload, 
it reduces the mean TTFT by 92\% and the mean TPOT by 24\% compared to vLLM, 
and it reduces the P99 TPOT by 13\% compared to llm-d---the second-best policy---with a similar TTFT.
The improved performance mainly comes from being {\kvcache}-aware
without sacrificing load balancing.
To examine {\sys}'s behavior,
{\fig{fig:eval-e2e-kvcache}} plots the {\kvcache} hit ratio for different systems
on the ChatBot workload.
We can see that {\sys} consistently achieves a {\kvcache} hit ratio comparable to that of other
{\kvcache}-aware policies, and its ratio is significantly higher than that of the
{\kvcache}-unaware policy (vLLM).
Meanwhile,
{\fig{fig:eval-e2e-imbalance}} further analyzes the imbalance
following the approach used in \textsection{\ref{sec:challenge-balancing}}.
For brevity, we compare {\sys} only with llm-d---the second-best policy on the ChatBot trace.
We can see that {\sys} achieves a better-balanced load than llm-d.

\begin{figure*}[!t]
    \centering
    \includegraphics[width=0.94\linewidth]{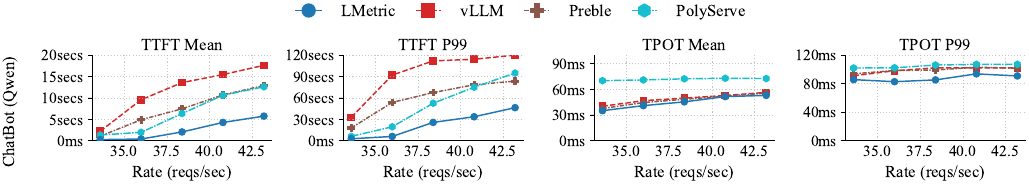}\\[-3pt]
    \caption{\small{
        End-to-end performance of {\sys} alongside two recent research baselines
        (Preble and PolyServe) under the same setting as {\fig{fig:scaling-test}}, with vLLM included as a reference.
        Model: Qwen3-30B on the ChatBot~(Qwen) workload.
        }}
    \label{fig:eval-academic-e2e}
\end{figure*}

\stitle{Performance under different request rates. \,}
{\fig{fig:scaling-test}} further shows how {\sys} performs under different request arrival rates.
The results are largely consistent with those under a fixed request rate setting:
{\sys} outperforms all baselines across different traces and request rates,
except for the ToolAgent trace,
where {\sys} exhibits a slightly higher (10\%) mean TTFT than llm-d
but achieves a 30\% lower TPOT.
This may be because a simulation-based approach can better estimate
the prefill workload than our simple P-token indicator.
Nevertheless, {\sys} still achieves the lowest TPOT without relying on a complex simulator,
as it considers both {\kvcache} management and (decode) load balancing.
The performance gap between baselines widens as the request rate increases,
because under high load, a more balanced and {\kvcache}-aware scheduling strategy
can yield higher overall system throughput,
leading to faster consumption of queued requests.

\subsection{Comparison with Recent Research Schedulers}
\label{sec:eval-academic}

\nospacestitle{Baselines. \,}
We further compare {\sys} with two state-of-the-art
research schedulers: Preble~\cite{DBLP:conf/iclr/SrivatsaHAL025} and PolyServe~\cite{DBLP:journals/corr/abs-2507-17769}.
As in the previous section, both baselines run in our
router framework (\textsection{\ref{sec:factory}}):
(1) Preble's open-source release runs on a different router stack from ours,
(2) PolyServe is not open-sourced,
and re-implementing both in our framework enables an apples-to-apples comparison.
We have carefully tuned different implementations and their hyperparameters to achieve 
the best performance, as detailed in \textsection{\ref{sec:appendix}}. \\[-15pt]

\begin{itemize}[leftmargin=*,leftmargin=10pt,itemindent=0pt]
\item \textbf{Preble~\cite{DBLP:conf/iclr/SrivatsaHAL025}} adopts a hybrid
    filter-based (\textsection{\ref{sec:filter-combination}})
    and linear-combination (\textsection{\ref{sec:linear-combination}})
    scheme.
    It first filters instances with high {\kvcache} hit ratios
    based on a threshold ($T=0.5$ tuned in \textsection{\ref{sec:appendix-compare-preble-v1}}). 
    If the filter returns any instance, 
    it routes to the one with the highest hit ratio; otherwise, 
    it routes the request to the instance with the highest linear-combined score over {\kvcache} and load indicators.
    \\[-15pt]

\item \textbf{PolyServe~\cite{DBLP:journals/corr/abs-2507-17769}} is a
    simulation-based scheduler
    (\textsection{\ref{sec:prediction-combination}}) that optimizes a
    different objective from the other baselines and {\sys}: meet the SLO
    while creating a load gradient across instances that facilitates
    auto-scaling. Using the simulator's predicted TTFT and TPOT, PolyServe
    routes each request to the most loaded instance whose predicted
    latency still meets the SLO bounds $\text{SLO}_{\text{TTFT}}$ and
    $\text{SLO}_{\text{TPOT}}$; when no instance is feasible, it falls
    back to the instance with the lowest predicted TPOT. 
    \\[-15pt]
\end{itemize}

\begin{figure}[!t]
    \centering
    \includegraphics[width=0.98\linewidth, trim=0.25cm 12.45cm 18.05cm 0.25cm, clip]{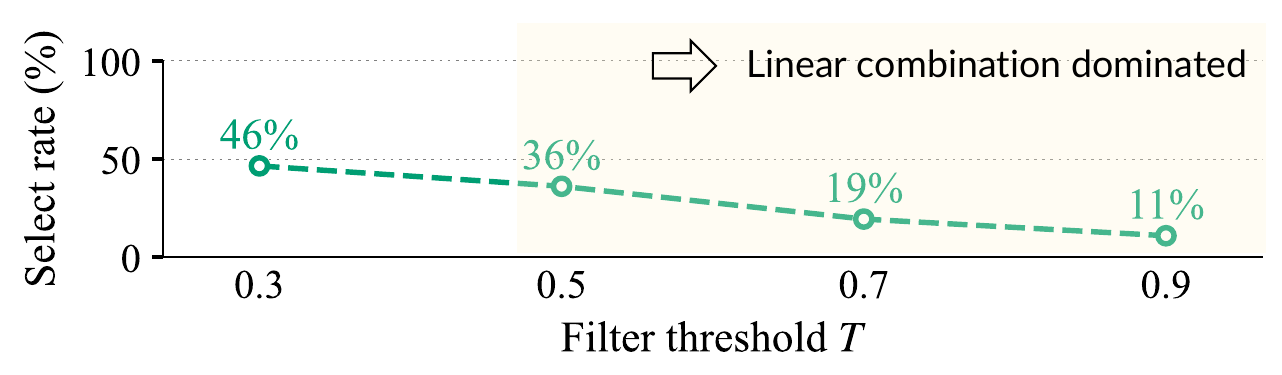} \\[-3pt]
    \caption{\small{
        Preble's {\kvcache}-aware branch
        selection rate under different filter thresholds $T$.
        }}
    \label{fig:eval-preble-threshold}
\end{figure}

\stitle{Overall performance under different request rates. \,}
{\fig{fig:eval-academic-e2e}} reports mean and P99 TTFT/TPOT for
{\sys} and the baselines on ChatBot~(Qwen) with Qwen3-30B.
Other workloads share similar results.

Compared with Preble, 
{\sys} reduces mean TTFT by 56\% and mean TPOT by 8\%,
and it reduces P99 TTFT by 45\% and P99 TPOT by 16\%.
Although Preble is faster than vLLM thanks to {\kvcache}-awareness,
it performs similarly to the linear-combination baselines that {\sys} outperforms,
because it falls back to the linear-combination branch most of the time
(see {\fig{fig:eval-preble-threshold}}).
Note that lowering the threshold to reduce
the linear-combination branch selection rate does not necessarily improve Preble's performance,
as it biases the scheduling towards {\kvcache} and thus sacrifices load balancing.

\begin{figure}[!t]
    \centering
    \includegraphics[width=0.94\linewidth]{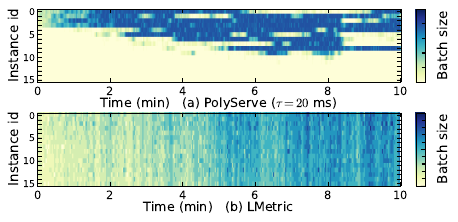} \\[-2pt]
    \caption{\small{%
        Running batch size across $16$ instances over a $10$-minute
        window, under PolyServe and {\sys}.
        Workload: ChatBot~(Qwen) trace at $18.75$\,reqs/sec on Qwen3-30B
        instances.
    }}
    \label{fig:polyserve-utilization-heatmap}
\end{figure}

Compared with PolyServe, {\sys} achieves lower mean and P99 TTFT and TPOT
across all request rates in {\fig{fig:eval-academic-e2e}}.
The gap reflects PolyServe's design objective:
instead of balancing load, PolyServe creates a load gradient across instances
so that idle ones can be released by auto-scaling.
Concentrating load this way raises instance utilization but degrades
per-request latency. 
{\fig{fig:polyserve-utilization-heatmap}} illustrates this trade-off:
PolyServe loads instances $0$--$8$ and leaves $9$--$15$ idle,
while {\sys} spreads the same aggregate workload across all $16$ instances.

\subsection{In-production Evaluation}
\label{sec:eval-product}

\begin{figure*}[!t]
    \centering
    \includegraphics[width=.96\textwidth, trim=0.25cm 19.25cm 5.0cm 0.25cm, clip]{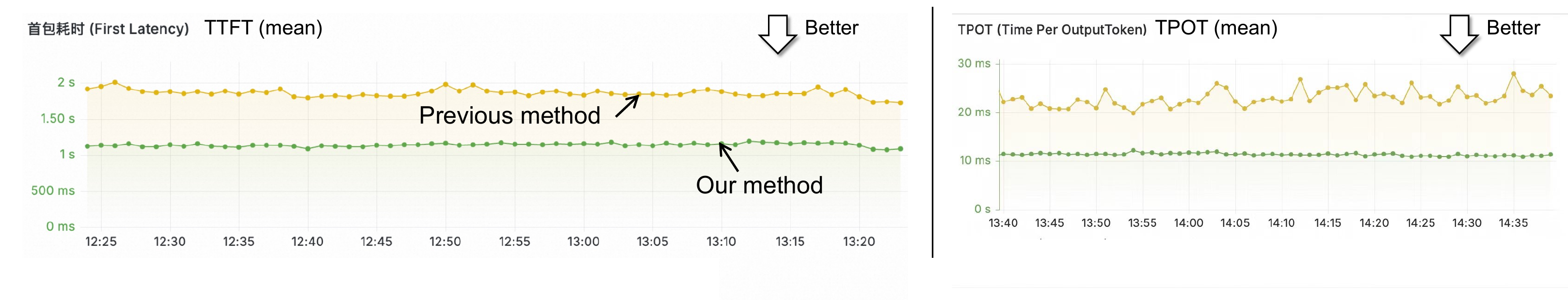} \\[-15pt]
    \caption{\small{Production deployment results of {\sys} 
    compared with the scheduler in {\company}.}}
    \label{fig:production-deployment}
\end{figure*}

\nospacestitle{Deployment overview and methodology to validate performance improvements. \,}
{\sys} now serves production traffic at {\company} as the scheduler for a
Qwen3.5-27B cluster, our first full-scale deployment.
To measure the impact on service quality, we report the observed performance
from a canary release.
On one day in May 2026, we split the traffic that previously fed one
production cluster across two clusters:
1/3 of the traffic went to a cluster running {\sys},
and the remaining 2/3 to a cluster running {\company}'s prior scheduling method.
To ensure a fair comparison, we sized the clusters so that
each received the same reqs/GPU.
The {\sys} cluster contained several hundred GPUs;
the detailed cluster setup is confidential and thus omitted.

\stitle{Performance. \,}
{\sys} reduces both TTFT and TPOT in production:
{\fig{fig:production-deployment}} compares the two clusters 
using screenshots of {\company}'s internal performance dashboard.
The snapshots show that, compared with {\company}'s prior scheduler,
mean TTFT and mean TPOT drop by 39\% and 51\% respectively.
The results imply that {\sys} can serve the same workload with fewer GPUs under the same SLO.

%% file: dislim.tex
\section{Discussion}
\label{sec:dislim}

\nospacestitle{Scheduling under PD-disaggregation. \,}
Scheduling under PD-disaggregation is different (and we argue simpler) than our
targeted PD-colocation, for two reasons.
First, routing decode requests primarily needs to consider load balancing.
Existing systems~\cite{ai-Dynamo,aibrix,llm-d,305212} already handle this effectively with simple indicators
like batch size~\cite{aibrix} or total tokens~\cite{ai-Dynamo}.
Second, routing prefill requests can address {\kvcache}-awareness and prefill load balancing
with a unified indicator---the number of new prefill tokens after cache hits.
Using this indicator as a scheduling score naturally combines both objectives
without explicit hyperparameter tuning~\cite{aibrix,ai-Dynamo},
and we find it effective in our analysis.
On the other hand, PD-disaggregation does introduce new scheduling challenges,
such as managing prefill and decode clusters with different capacities,
which we leave to future work.

\stitle{Scheduling under heterogeneous deployments. \, }
In real-world production, model services are heterogeneous in both models and
GPU types, while {\sys} targets scheduling for a single model under homogeneous GPUs.
Nevertheless, {\sys} still works in this setting because providers, to cope with
this complexity, logically partition their deployment into multiple clusters,
each serving a single model on a single GPU type.
The global scheduler inside each cluster therefore sees a homogeneous pool of
instances, where {\sys} applies.

\stitle{KVCache sharing. \,}
Some serving clusters support {\kvcache} sharing: an instance can fetch the
requested {\kvcache} from another instance through RDMA when it does not hold
the cache locally~\cite{305212}. This mechanism reduces the cost of scheduling a request to an
instance without {\kvcache}, but it does not remove the benefits of
{\kvcache}-aware scheduling provided by {\sys} because local {\kvcache} hits are still cheaper than
remote {\kvcache} fetches in two ways. First, a local hit avoids a remote
transfer on the TTFT path. Second, it avoids keeping the same {\kvcache} on two
instances, which saves memory and reduces cache pressure.

%% file: related.tex
\section{Related Work}
\label{sec:related}

\nospacestitle{LLM requests global scheduling. \,}
{\sys} continues the line of research on scheduling LLM requests in a cluster~\cite{vllm,DBLP:journals/corr/abs-2506-05871,DBLP:conf/iclr/SrivatsaHAL025,
305212,DBLP:journals/corr/abs-2507-17769,llm-d,DBLP:journals/corr/abs-2602-06502}.
To the best of our knowledge, all these methods target both
{\kvcache}-awareness and load balancing
by combining indicators for each objective through the three approaches
discussed in \textsection{\ref{sec:principles}}.
{\sys} reuses these indicators but combines them with a simple yet efficient
multiplication combinator.

\stitle{LLM requests scheduling within an instance. \,} 
These works are orthogonal to {\sys}'s global scheduling:
they optimize request execution within an instance.
Sarathi-Serve~\cite{298679} introduces chunked prefill, which splits long prefill requests into smaller chunks
to reduce stalls for co-located decode requests.
VTC~\cite{DBLP:conf/osdi/0007CLZ0ZGS24} adopts token-based admission control
to achieve fairness.
FairBatching~\cite{DBLP:journals/corr/abs-2510-14392} uses a linear-time analytical model
to prioritize prefill versus decode tokens and dynamically form batches.
A good global scheduler can further improve these local schedulers,
e.g., by reducing overloads that are difficult to handle locally.

\stitle{Optimizing LLM serving. \,}
Beyond scheduling, LLM serving systems improve performance by increasing
{\kvcache} hit rates~\cite{305212,10.5555/3768039.3768067},
adding model-serving elasticity~\cite{DBLP:conf/osdi/ZhangWLWS0025,DBLP:conf/sosp/Xiang0QYZYZL0025},
and improving GPU execution efficiency~\cite{vllm,DBLP:conf/asplos/KamathPM0RP25,flashinfer,DBLP:conf/iclr/Dao24},
just to name a few.
These techniques coexist with scheduling optimizations.

%% file: concl.tex
\section{Conclusion}
\label{sec:concl}

\noindent
We contribute the 
first multiplicative combinator for high-quality LLM request scheduling,
achieving both {\kvcache}-awareness and load balancing in a hyperparameter-free manner.
Evaluations on real-world workloads covering chatbots and agents 
confirm the benefits of our approach, and our method has been deployed in production 
with confirmed effectiveness.

%% file: ack.tex
\section{Acknowledgment}
\noindent
We sincerely thank the OSDI'26 reviewers 
for their insightful comments
and Neeraja J. Yadwadkar for shepherding our paper.
We also thank Jinyu Gu for valuable feedback and for suggesting
the classic system term ``simple is better'' for our paper.
This work was supported in part by
the National Natural Science Foundation of China (No.~62572302),
the Fundamental and Interdisciplinary Disciplines Breakthrough Plan
of the Ministry of Education of China (No.~JYB2025XDXM122),
and the Alibaba Innovative Research Program.

%% file: appendix-v1.tex
\section{Appendix}
\label{sec:appendix}

\subsection{More on Preble}
\label{sec:appendix-compare-preble-v1}
\begin{center}
    \includegraphics[width=0.9\linewidth, trim=0.25cm 18.9cm 37.65cm 0.25cm, clip]{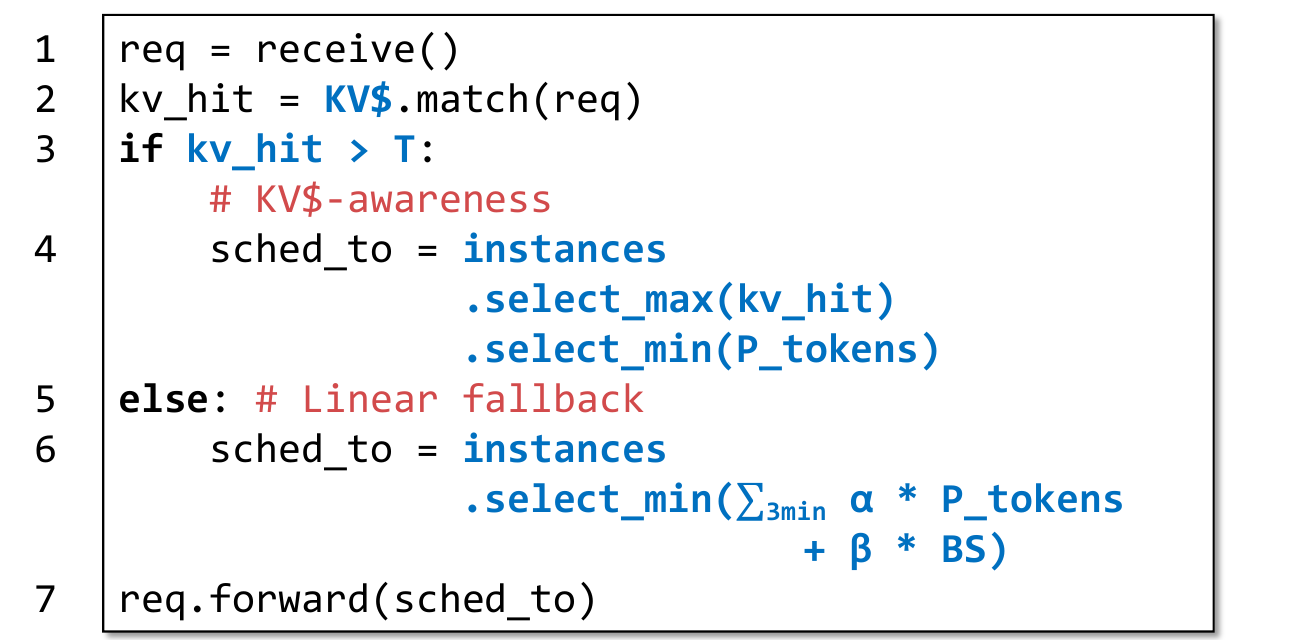} \\[5pt]
    \captionof{figure}{\small{%
        The pseudocode of Preble's method. 
    }}
    \label{fig:preble-appendix}
\end{center}

\nospacestitle{Preble method in detail.\,}
Preble~\cite{DBLP:conf/iclr/SrivatsaHAL025} adopts
two of the combination methods we identify
in a hybrid form: a {\kvcache}-aware filter
(\textsection{\ref{sec:filter-combination}}) on top of a
linear-combination fallback
(\textsection{\ref{sec:linear-combination}}).
\fig{fig:preble-appendix} shows the pseudocode using our 
scheduling language.
After receiving a request,
Preble first filters the instances with high {\kvcache} hits---i.e., 
instances whose cached prefix covers
more than a threshold $T$ of the prompt (line 3).
Among the instances tied for the highest {\kvcache} hit ratio,
it routes the request to the one with the least prefill load (line 4).
Otherwise, it uses a linear-combination score over all instances and
selects the one with the smallest score (line 6).

More specifically,
Preble adopts the following scoring function for the fallback path:

\begin{equation*}
    \arg\min_i \sum_{r \in W_i} \big(\text{PT}_r + \text{DT}_r\big),
\end{equation*}
\noindent
a sum of per-request prefill time $\text{PT}_r$ and decode time
$\text{DT}_r$ over recent requests $W_i$ routed to instance $i$.
Preble realizes $\text{PT}_r$ by assigning a
pre-determined per-token prefill cost to each newly prefilled
token of $r$, and $\text{DT}_r$ by assigning a pre-determined
per-request decode cost to $r$ itself, both aggregated per
instance over a 3-minute sliding window.
In its implementation~\cite{aibrix-prefix-cache-preble}, 
Preble derives these two costs from
exact indicators exported by the engine, see below. 

\begin{equation*}
    \arg\min_i \;
    \Big(\,
        \alpha \cdot \!\!\sum_{\text{window}}\! \text{P-token}_i
        \;+\;
        \beta \cdot \!\!\sum_{\text{window}}\! \text{BS}_i
    \,\Big).
\end{equation*}

The fallback score of Preble is therefore a variant of the
linear combination analyzed in
\textsection{\ref{sec:linear-combination}}.

\stitle{Setup and tuning. \,}
We re-implement Preble inside {\sys}'s router framework 
for two reasons: (1) to enable an apples-to-apples comparison with our method,
and (2) the open-source reference does not sustain our request rate, an issue tracked upstream~\cite{aibrix-issue-1580}.

We carefully tune the configurations of Preble to ensure a fair comparison.
Preble has three knobs to tune\footnote{\footnotesize{%
The fallback score actually depends only on the ratio $\alpha/\beta$, so the two
coefficients carry one degree of freedom. Nevertheless, Preble's implementation exposes
$\alpha$ and $\beta$ as separate parameters, so we tune both to match it.}}:
the filter threshold $T$ and the linear-combination coefficients.
Tuning the combination space is expensive,
so we focus on $T$ and fix the linear-combination coefficients $\alpha$ and $\beta$
using the profiling method described in Preble's paper.
Surprisingly, we find $T$ has little impact on
performance (see {\fig{fig:preble-threshold-sweep}}),
and Preble's default ($T = 0.5$) is already optimal on our traces.
As a result, all experiments in \textsection{\ref{sec:eval-academic}}
use the published default $T = 0.5$.

\begin{center}
    \includegraphics[width=0.98\linewidth]{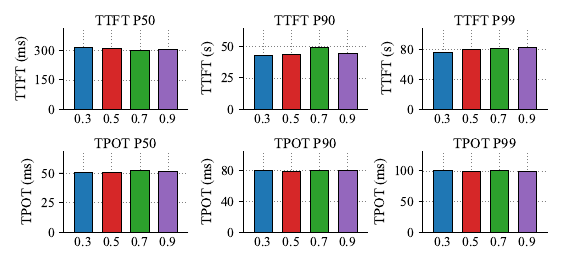} \\[-2pt]
    \captionof{figure}{\small{%
        Performance of Preble as the filter threshold $T$ varies
        on the ChatBot~(Qwen) trace; model: Qwen3-30B.
    }}
    \label{fig:preble-threshold-sweep}
\end{center}

\begin{center}
    \includegraphics[width=0.98\linewidth]{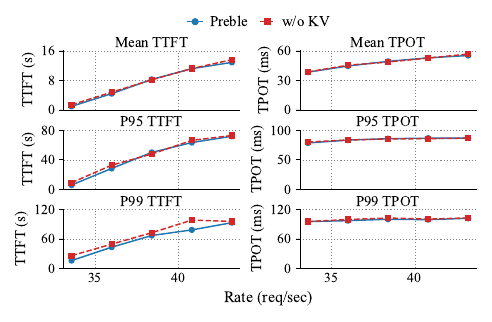} \\[-10pt]
    \captionof{figure}{\small{%
        Comparison of Preble with and without the {\kvcache}-aware filter.
    }}
    \label{fig:preble-no-kvcache-indicator}
\end{center}

\stitle{Comparison of Preble with and without the {\kvcache}-aware filter. \,}
To characterize how much of Preble's behavior comes from its
{\kvcache}-aware filter on top of the linear-combination fallback
(\textsection{\ref{sec:linear-combination}}), we compare the default
($T=0.5$) against $T=1$, which disables the filter and routes purely by the
fallback score. \fig{fig:preble-no-kvcache-indicator} reports the comparison
on the ChatBot~(Qwen) trace with $16$ Qwen3-30B instances. We can see that
the filter yields a measurable improvement, but the bulk of Preble's
performance on this trace is dominated by its linear-combination
component, consistent with the $T$ sweep in
\fig{fig:preble-threshold-sweep} and our analysis in
\textsection{\ref{sec:eval-academic}} (\fig{fig:eval-preble-threshold}).

\subsection{More on PolyServe}
\label{sec:appendix-compare-polyserve-v1}

\begin{figure}[th]
    \centering
    \includegraphics[width=0.9\linewidth, trim=0.25cm 18.79cm 37.56cm 0.25cm, clip]{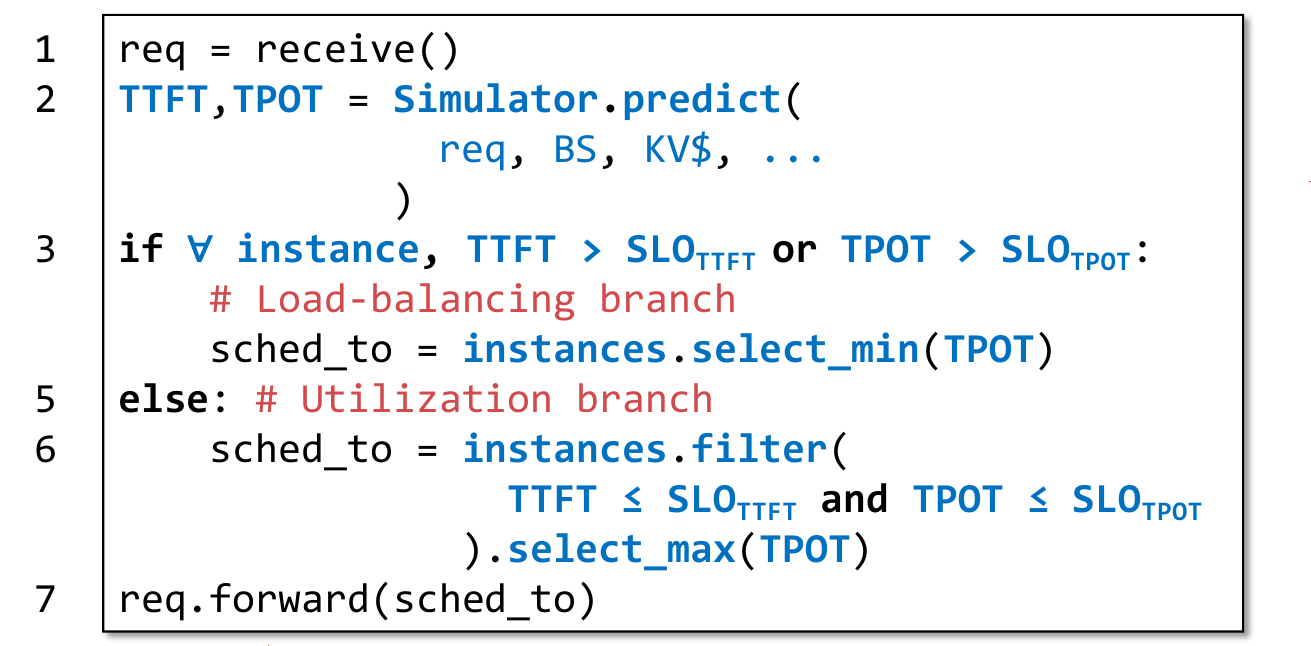} \\[5pt]
    \caption{\small{%
        The pseudocode of PolyServe's simulation-based filter
        scheduler.
    }}
    \label{fig:polyserve-appendix}
\end{figure}

\nospacestitle{PolyServe method in detail.\,}
PolyServe~\cite{DBLP:journals/corr/abs-2507-17769} is a simulator-based
scheduler that optimizes a different objective: meet the SLO while creating a
load gradient across instances that facilitates auto-scaling.
\fig{fig:polyserve-appendix} shows its scheduler in our scheduling language. On
each request, PolyServe first calls the simulator of
\textsection{\ref{sec:prediction-combination}} to predict per-instance TTFT and
TPOT under the new request, conditioned on per-instance batch size
\textbf{BS} and {\kvcache} footprint (line 2). It then branches on the
simulator's output. If any instance meets the SLO bounds
$\text{SLO}_{\text{TTFT}}$ and $\text{SLO}_{\text{TPOT}}$ (line 5), PolyServe
takes the utilization branch: filter to feasible instances and route to the
one with the highest predicted TPOT, i.e., the most loaded feasible instance
(line 6). Otherwise (line 3), PolyServe falls back to load balancing and
routes to the instance with the lowest predicted TPOT (line 4).

\begin{figure}[th]
    \centering
    \includegraphics[width=0.98\linewidth]{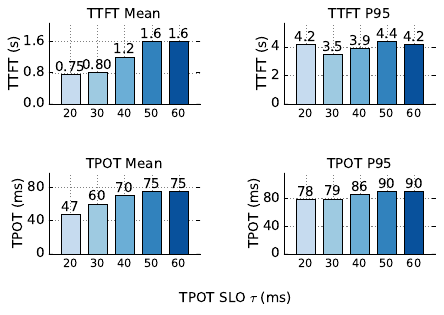} \\[-2pt]
    \caption{\small{%
        PolyServe end-to-end TTFT and TPOT under different
        TPOT-SLO thresholds $\tau$ on the ChatBot~(Qwen) trace
        at $35.0$\,reqs/sec on the same $16$ Qwen3-30B
        instances as
        \textsection{\ref{sec:eval-academic}}.
    }}
    \label{fig:polyserve-tau-sweep}
\end{figure}

\stitle{Setup and tuning. \,}
PolyServe's scheduling is parameterized by the SLO bounds
$\text{SLO}_{\text{TTFT}}$ and $\text{SLO}_{\text{TPOT}}$, so its performance
depends on the configured SLO. Following PolyServe's
paper~\cite{DBLP:journals/corr/abs-2507-17769}, we tune the SLO and adopt the
best-performing setting in \textsection{\ref{sec:eval-academic}}. \fig{fig:polyserve-tau-sweep} reports the tuning of $\tau$
(denoting $\text{SLO}_{\text{TPOT}}$) on the ChatBot~(Qwen) trace, with
$\text{SLO}_{\text{TTFT}}$ held fixed; we adopt $\tau = 20$\,ms.
We show only this $\tau$ sweep because $\text{SLO}_{\text{TTFT}}$
had little impact on end-to-end performance on our traces.